\documentclass[preprint2,numberedappendix,iop]{emulateapj-rtx4}
\usepackage{graphicx,graphics,amsmath}
\usepackage{natbib}
\usepackage{bm,url}
\usepackage{times}
\usepackage{amssymb}
\usepackage{xcolor}
\usepackage{ulem}

\definecolor{orchid}{RGB}{163,73,164} % Melinda

\usepackage{float}
\usepackage{arydshln}
\graphicspath{{./fig/}{./png/}}

\def\red{\textcolor{red}}

\def\bl{Babcock--Leighton}
\def\2by2{$2\times2$ D}
\newcommand{\Fig}[1]{Figure~\ref{#1}}

\newcommand{\Tab}[1]{Table~\ref{#1}}

\newcommand{\mps}{m~s$^{-1}$}
\newcommand{\cmss}{cm$^2$~s$^{-1}$}

\begin{document}
%\title{Correlation between the polar magnetic field and the following 
%cycle from observations and Babcock--Leighton dynmo models}
\title{The polar precursor method for solar cycle prediction: 
comparison of predictors and their temporal range}
\medskip
\author{Pawan Kumar$^1$, Melinda Nagy$^2$, Alexandre Lemerle$^{3,4}$, 
Bidya Binay Karak$^1$,  and Kristof Petrovay$^2$}
\email{karak.phy@iitbhu.ac.in}
\affiliation{$^1$Department of Physics, Indian Institute of Technology 
(Banaras Hindu University), Varanasi, India\\
$^2$E\"otv\"os Lor\'and University, Department of Astronomy, Budapest,
Hungary\\
$^3$Coll\`ege de Bois-de-Boulogne, Montr\'eal, QC, Canada\\
$^4$D\'epartement de Physique, Universit\'e de Montr\'eal, Montr\'eal, QC, 
Canada
}
%\correspondingauthor{Bidya Binay Karak}
%\email{karak.phy@iitbhu.ac.in}

\date{\today}

\begin{abstract}
The polar precursor method is widely considered to be the most robust physically motivated method 
to predict the amplitude of an upcoming solar cycle.
It uses indicators of the magnetic field concentrated near the 
poles around sunspot minimum. 
Here, we present an extensive performance analysis of
various such predictors, based on both observational data 
(WSO magnetograms, MWO polar faculae counts and Pulkovo 
$A(t)$ index)
and outputs (polar cap magnetic flux and global dipole moment) 
of various existing 
flux transport 
dynamo models.
We calculate Pearson correlation coefficients ($r$) of the predictors with the 
next cycle amplitude as a function of time measured from several solar 
cycle landmarks:
setting
$r= 0.8$ as a lower limit for acceptable predictions, we find
that observations and models alike indicate that the earliest time
when the polar predictor can be safely used is 4 years after polar
field reversal. 
This is typically 2--3 years before solar minimum and
about 7~years before the predicted maximum, considerably extending the {usual}
temporal scope of the polar precursor method. 
Re-evaluating the predictors another 3 years later, at the time of solar minimum, further increases the correlation level to $r\ga 0.9$. 
As an illustration of the result, we determine the predicted amplitude of Cycle 25 based on the value of the WSO polar field at the now official minimum date of December 2019 as $126\pm 3$. 
A forecast based on the value in early 2017, 4~years after polar reversal would have only differed from this final prediction by 
$3.1\pm 14.7$\%.

\end{abstract}
%\maketitle

%____________________________________________________________

\section{Introduction}
\label{sec:int}

The Sun's magnetic field increases and decreases in time with a
polarity reversal every 11 years. This 
cycle of magnetic field is widely studied using the sunspot number and areas as they are the
signatures of the strong field and are available for a longer duration
than the direct observations of the surface magnetic field, which are  available in the form of synoptic maps only since around 1955s. 
The strength of the solar magnetic field, however,
varies from
cycle to cycle in an irregular manner \citep{Hat15}. The
Maunder minimum, an extended period of weaker field, is an extreme
example of such irregular behaviour. It is this variable magnetic
field that drives the 
fluctuations 
of the solar wind and 
{terrestrial} 
space weather, which may have hazardous effects on {human activities that depend on the stability of our magnetospheric environment (space missions, satellites, telecommunications, etc.)}.
{To assess and prevent those risks,} 
reliable predictions of future solar activity are now more than ever essential\footnote
{See, e.g., \url{https://www.swpc.noaa.gov/products/solar-cycle-progression}~.}.

Many attempts have been made to predict an upcoming solar cycle based on the extrapolation of the time series of some proxy of solar activity. However, most of the time, they produce diverging results \citep{Pesnell12}.

Predictions of the solar cycle based on dynamo theory have also been made \citep{Petrovay20}.
In most dynamo models, the polar magnetic field is a natural predictor of the following cycle. Indeed, there is considerable evidence that the solar dynamo is a magnetohydrodynamical $\alpha\omega$-dynamo \citep{Charbonneau:LRSP2} where the strong toroidal field, manifest in the form of sunspots, is generated by the winding up of a weaker poloidal magnetic field by differential rotation. This windup is essentially a linear process; hence the amplitude of a solar cycle is expected to be proportional to the amplitude of the poloidal field around the start of the cycle. As this 
{poloidal}
field is strongly concentrated to the poles, the amplitude of the polar 
{surface}
magnetic field is a plausible measure of its amplitude.

Following this idea, \citet{Sch78} 
used the polar field strength at the solar minimum to make the first
prediction of the sunspot number during Cycle 21. Later this idea was
validated by many authors using different proxies of the polar field,
such as $aa$-index, polar faculae, and active networks 
\citep{MMS89,CCJ07, JCC07, WS09, KO11, Muno13, Priy14}. 
Furthermore, surface flux transport models, which provide the evolution of the surface radial field by utilizing the observed BMRs and large-scale flows 
have also been used to predict the amplitude of the upcoming solar cycle 
\citep{Iijima17, jiang2018predictability, UH18}.

As a whole,
predictions based on these polar precursors tend to yield much more
consistent results than time series methods, and there is now wide
consensus that this is the most robust physically motivated method to
predict the amplitude of an upcoming solar cycle.

However, when applying polar precursor methods, a variety of choices
need to be made regarding 
the measures or proxies of the
poloidal field to use, and 
{the exact time at which}
%at exactly what time 
they are evaluated. The
purpose of the present work is to make a systematic analysis of the
performance of various such predictors
{and determine a temporal window when their evaluation is optimal}. 
Some of the questions we
address are:  Is it the polar field at the time of its peak value that best
determines the strength of the next cycle or is it 
%the field 
at a different time? Is there 
an acceptable 
window over which we can use the polar
field data to make a prediction? Do we need to wait until the activity 
minimum to make a reliable prediction of the next cycle?

\begin{table*}
\centering
\caption{Relative time shifts (in years) between different solar cycle landmarks: observations and models.
}
\begin{tabular}{lccccccc}
&\multicolumn{4}{c}{Observations} \\
\cline{1-8}
Shift between \\
\hline
$t_{{\rm SSA,min} ,i}$ to $t_{{\rm SSA,min},i+1}$ & 10.83$\pm$0.83 \\ %\hline
$t_{{\rm SSA,min} ,i}$ to $t_{{\rm SSA,max},i}$ & 4.58$\pm$0.81   \\
\hline                             &
\multicolumn{1}{c}{Faculae}  &&  \multicolumn{1}{c}{PF} &&  \multicolumn{1}{c}{DM}  &&\multicolumn{1}{c}{$A(t)$ index} \\
%\hline
\cline{2-8}
$t_{{\rm SSA,min} ,i}$ to $t_{{\rm p,rev},i}$    &
5.01$\pm$0.64   &&  4.12$\pm$0.36   && 3.53$\pm$0.27  && 3.55$\pm$0.48  \\
%\hline
$t_{{\rm SSA,max} ,i}$ to $t_{{\rm p,max},i}$ &
5.93$\pm$1.89   &&  4.40$\pm$0.99   &&  1.84$\pm$0.80   &&  4.85$\pm$1.47   \\
%\hline
$t_{{\rm p,max},i-1}$ to $t_{{\rm SSA,max},i}$ &
4.64$\pm$1.18    && 5.78$\pm$1.89     &&  7.88$\pm$2.43    &&  5.66$\pm$1.26     \\
%\hline
$t_{{\rm p,rev},i}$ to $t_{{\rm p,max},i}$ & 
5.76$\pm$1.49    &&   5.48$\pm$0.74   && 3.52$\pm$1.16     &&  4.96$\pm$0.78   \\
%\hline
$t_{{\rm p,rev},i}$ to $t_{{\rm SSA,min},i+1}$ & 
5.60$\pm$1.01    &&  6.81$\pm$1.53    && 7.39$\pm$1.34   &&  6.88$\pm$0.61    \\
%\hline
$t_{{\rm p,max},i-1}$ to $t_{{\rm SSA,min},i+1}$   & 
10.72$\pm$1.45   &&  10.94$\pm$2.28    && 12.92$\pm$3.42   &&  12.08$\pm$1.28  \\
\hline
\end{tabular}
%\label{tab:obsstat}
%\end{table*}
\centering
\begin{tabular}{lcccc ccccc}
&\multicolumn{6}{c}{Models} \\
\cline{1-10}
 &\multicolumn{1}{c}{2DR1}    && \multicolumn{2}{c}{3DR1}            && \multicolumn{2}{c}{2$\times$2D-R1} \\
\cline{2-3}
\cline{5-6}
\cline{8-9}
\cline{10-10}
%\hline
Shift between  &    PF  & DM && PF    & DM   &&  PF && DM   \\
\hline
$t_{{\rm SSN,min} ,i}$ to $t_{{\rm SSN,min},i+1}$ &  11.12$\pm$1.26 & $---$      &&  10.50$\pm$0.74  & $---$    &&  10.25$\pm$0.93 && $---$ \\
%\cline{1-7}
$t_{{\rm SSN,min} ,i}$ to $t_{{\rm SSN,max},i}$ &   5.76$\pm$0.53  & $---$  &&  5.29$\pm$0.47  & $---$    &&   $4.92\pm$0.83 && $---$  \\
%\cline{1-7}
$t_{{\rm SSN,min} ,i}$ to $t_{{\rm p,rev},i}$    &   4.36$\pm$0.41 & 4.96$\pm$0.52     &&    5.52$\pm$0.47   & 4.54$\pm$1.62  && {3.10$\pm$0.72} &&  2.67$\pm$0.67 \\
%\cline{1-7}
$t_{{\rm SSN,max} ,i}$ to $t_{{\rm p,max},i}$ & 3.98$\pm$0.94 & 4.78$\pm$0.96    &&  5.40$\pm$1.10  & 3.71$\pm$1.37  && {2.57$\pm$1.52}  && 2.47$\pm$1.28 \\
%\cline{1-7}
$t_{{\rm p,max},i-1}$ to $t_{{\rm SSN,max},i}$ & 7.14$\pm$1.26 & 6.44$\pm$1.24     && 5.08$\pm$0.99    & 6.79$\pm$1.26 && {7.70$\pm$1.52} && 7.79$\pm$1.29 \\
%\cline{1-7}
$t_{{\rm p,rev},i}$ to $t_{{\rm p,max},i}$ & 5.74$\pm$1.14 & 5.62$\pm$0.98     && 5.31$\pm$1.05   &  4.49$\pm$1.47 && {4.38$\pm$1.33}  && 4.72$\pm$1.12 \\
%\cline{1-7}
$t_{{\rm p,rev},i}$ to $t_{{\rm SSN,min},i+1}$ & 6.75$\pm$1.03  & 6.26$\pm$0.93    && 4.98$\pm$0.55  & 5.98$\pm$1.30 && {7.14$\pm$0.92}  && 7.58$\pm$0.73  \\

$t_{{\rm p,max},i-1}$ to $t_{{\rm SSN,min},i+1}$  & 
%12.47$\pm$1.92   & 11.86$\pm$1.89   &&  10.16$\pm$1.09    & 12.05$\pm$1.41   &&  2.83$\pm$ 1.46&&  2.85$\pm$1.13 \\
12.47$\pm$1.92   & 11.86$\pm$1.89   &&  10.16$\pm$1.09    & 12.05$\pm$1.41   &&   13.02 $\pm$ 1.44&&  13.10$\pm$1.24 \\
\hline
\end{tabular}
\label{tab:obsstat}
\tablecomments{The number of cycles used in the analysis  are 270, 200, and 260, respectively for models 2DR1, 3DR1, and 2$\times$2D-R1. 
The symbol '$---$' indicates that the values are the same as given in the previous column.
}
\end{table*}

In Section 2 these questions are considered on the basis of the
available observational data. This analysis leads to  some interesting
findings, the most important of which is that the temporal scope of
the polar precursor method is longer than generally thought:
reliable predictions can be made already 4 years after the reversal of
the Sun's polar magnetic field, that is on average 3 years before the
minimum and 7 years before the next maximum. This significant
extension of the temporal scope of the polar precursor method is
potentially highly relevant for solar cycle prediction efforts. 

As, however, it is based on observational data covering only a limited
number of solar cycles, the statistical robustness of this result may
reveal somewhat doubtful. Hence, in Section 3 we extend our analysis to
some dynamo models that have been calibrated to represent reasonably
well the observed spatiotemporal variation of solar activity. As in
these models a higher number of cycles can be simulated, in this case
the size of the statistical sample is not a concern. On the other
hand, the three dynamo models considered are different in their
construction and in the parameter range where they operate, so a
direct comparison of the models with each other or with the real Sun
is questionable. Hence, we focus here only on those results where all
models agree with each other and the observations, and consider these
as robust and independent of model details.

As we conclude in Section 4, the possibility to predict an upcoming cycle on the basis of a
measure of the poloidal field 4 years after the observed time of polar
reversal is found to be one such robust result. To illustrate the result, in the Conclusions we also compare predictions of Cycle 25 that could have been made 4 years after polar reversal, i.e. in early 2017, to those based on the polar field amplitude at the now official date (December 2019) for the start of the new cycle.

Throughout the paper, we will use the generic symbolic notation $P$ for the different predictors (poloidal field measures) considered, and $T$ for the predicted variable characterizing the solar activity level (e.g. sunspot area, sunspot number or toroidal field amplitude).

%%%%%%%%%%%%%%%%%%%%%%%%%%%%%%%%%%%%%%%%%%%%%%%%%%%%%%%%%%%%%%%%%%

\section{Observational data analysis}
\label{sec:data}

\subsection{Data}
For observational measures and proxies of the solar poloidal field $P$, 
%in the present study 
we consider polar field strength and global dipole moment values derived from Wilcox Solar Observatory (WSO) magnetic field
measurements for the last four cycles (1974--2019); Mt. Wilson Observatory (MWO) polar 
faculae counts for the last ten cycles (1907--2011); and Pulkovo Observatory $A(t)$ index
series for the period 1915--1999.

The monthly time series of WSO polar magnetic field data\footnote
{\url{http://wso.stanford.edu/Polar.html}}, 
smoothed using a 
Gaussian filter with FWHM =  6 months \citep{Hat02}, is denoted here by $B_N(t)$ and $B_S(t)$ for the Northern and Southern hemispheres, respectively. As a further possible predictor we also consider the global solar dipole moment defined as
\begin{equation}                       
    {\rm DM} (t) = \frac32 \int_{-\pi/2}^{\pi/2} 
    B(\lambda,t)\sin\lambda\cos\lambda\, \mathrm{d}\lambda ,
 \label{eq:dipmom}
\end{equation}
where $\lambda$ is the heliographic latitude and $B=\frac 1{2\pi}\int_0^{2\pi} B_r \,d\phi$ is the azimuthally
averaged field strength. Values of ${\rm DM}(t)$ computed using this formula with WSO data for the period 1976--2017 were kindly provided to us by Jie Jiang \citep{jiang2018predictability}. 

The polar faculae 
count is considered a proxy of the polar field as they
share a significant correlation 
%with the polar field 
\citep{S91}. The MWO faculae 
counts used here were determined by \citet{muno12}\footnote
{\url{https://doi.org/10.7910/DVN/KF96B2}}.

The $A(t)$ index, representing the sum of the intensities of dipole
and octupole components of the Sun's large-scale magnetic field, as
reconstructed on the basis of Pulkovo H$_\alpha$ synoptic maps, is
obtained from \citet{Makarov01}. Since this index does not include the
quadrupolar component, symmetric to the equator, it is expected to scale
roughly with the antisymmetric part of the polar fields,
$(B_N-B_S)/2$.

To characterize the amplitude of solar cycles, we use the peak of the
sunspot area (SSA) distribution obtained from the Greenwich Royal 
Observatory\footnote
{\url{https://solarscience.msfc.nasa.gov/greenwch.shtml}},
%To smooth the data, we apply
smoothed using a Gaussian filter of FWHM $= 12$ months.

\subsection{Solar cycle landmarks}

{We consider two types of}
landmarks during a solar cycle to which the time 
{of evaluation of} 
the precursor $P$ 
%is evaluated 
can be bound: 
%we consider 
the maxima (or minima) of the smoothed sunspot area (SSA)\footnote
{\url{http://sidc.oma.be/silso/DATA/SN\_ms\_tot\_V2.0.txt}}, 
{from which we measure backward delays,} 
and the time of reversal (zero crossing) of the various poloidal field indicators $P$,
{from which we measure forward delays.} 
The relative time shifts between these landmarks varies from cycle to cycle; 
their average values are collected in Table~\ref{tab:obsstat} for later reference.
We note that in the polar faculae data, the epochs of polar field reversal is not available, and thus we take these as the same as the time of minima of the faculae.

\begin{figure}
\centering
\includegraphics[scale=0.35]{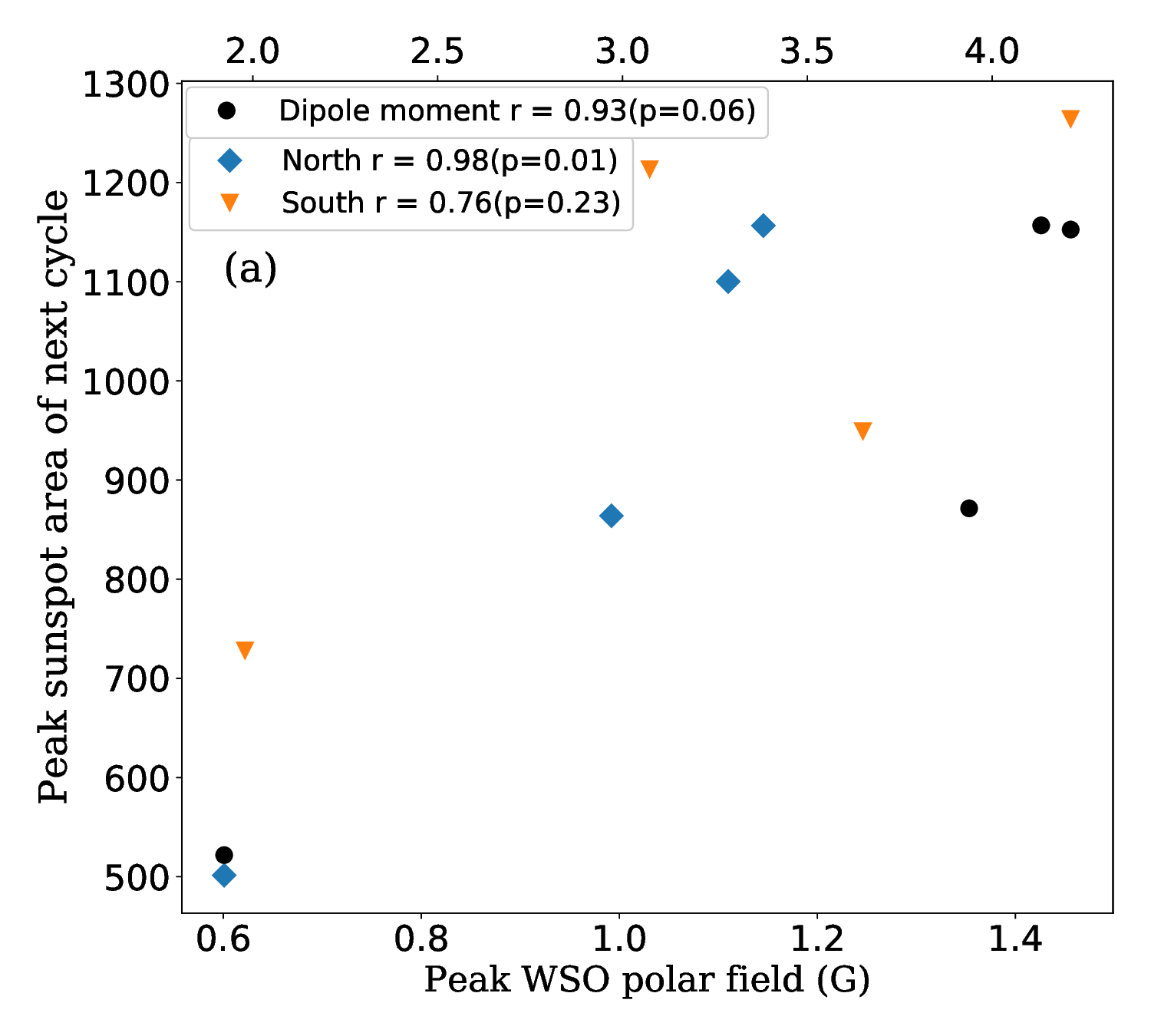}
\includegraphics[scale=0.332]{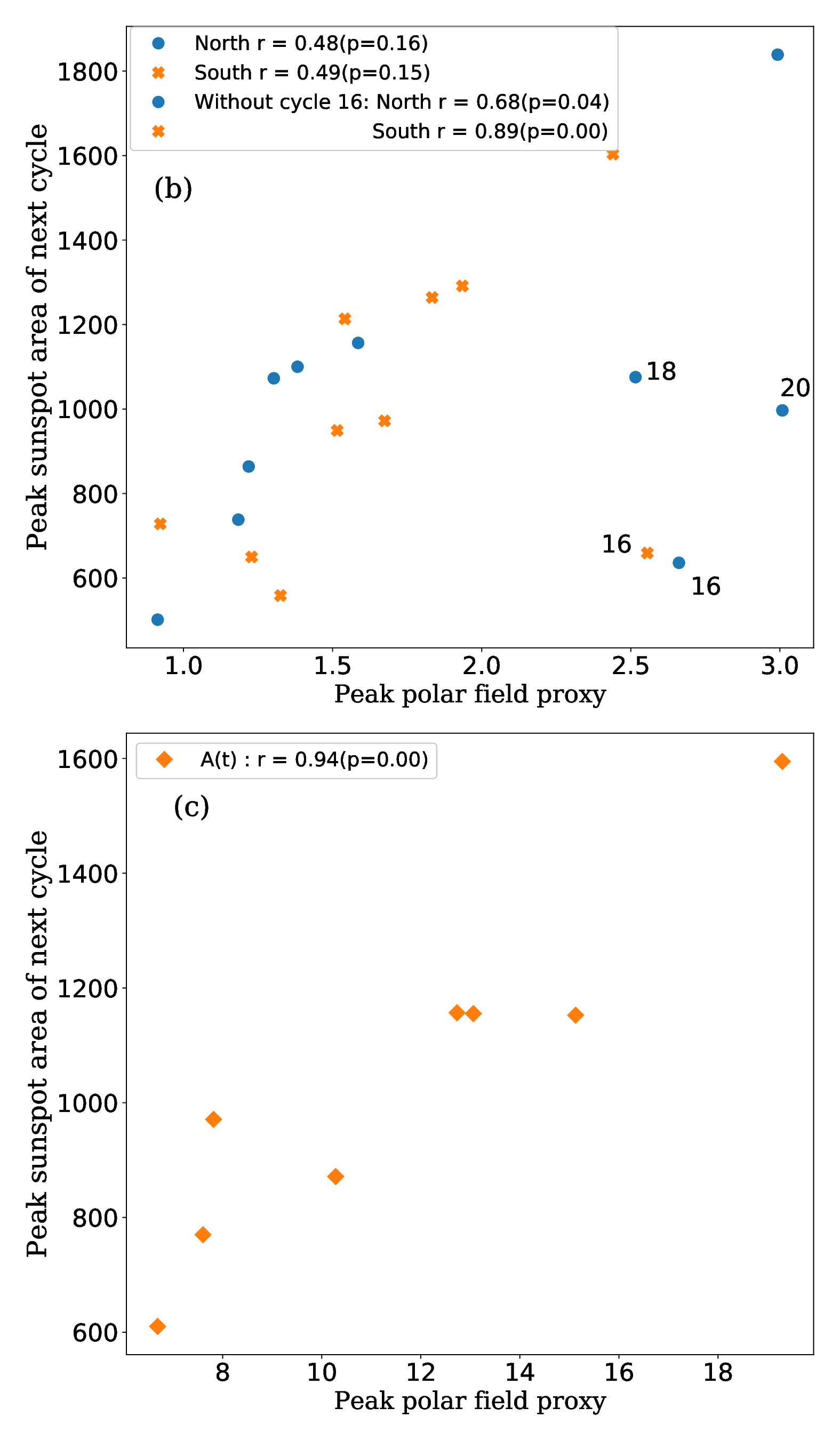}
\caption{Scatter plot of the peak value of the polar field (or its proxy) vs the peak sunspot area in the following cycle for 
(a) WSO polar field and dipole moment {(Cycles 21 to 24)}, 
(b) {MWO} polar faculae {(Cycles 14 to 23)}, and (c) {Pulkovo Observatory} $A(t)$ index {(Cycles 15 to 22)}. Sunspot areas in (a) and (b) are hemispheric peak values, while in (c) the peak value of the hemispheric average.
}
\label{fig:obscorr}
\end{figure}

\subsection{Correlations in the observed data}
\label{sec:obs}
Scatter plots of the peak values of various measures or proxies of the 
{poloidal field $P$}
%polar field 
against the amplitude of the following
sunspot cycle are shown in \Fig{fig:obscorr}. 
We note that Pearson correlations are given here and throughout this paper.

While the correlation is high in the WSO data (\Fig{fig:obscorr}(a)), it is less impressive in faculae data (\Fig{fig:obscorr}(b)). Particularly, in the
northern hemisphere, the correlation is relatively poor because of Cycles 16, 18, and 20. Note that faculae counts for Cycle 16 may be problematic, as already noted by \cite{Priy14}.
If we exclude only Cycle 16, then the correlation considerably improves.
\Fig{fig:obscorr}(c) shows that the $A(t)$ index, available for the longest time interval of all the considered polar field measures, performs quite well as a predictor. This is somewhat surprising, given that the index is based on a very rough reconstruction of the large scale solar magnetic fields: a simple two-valued function ($+1$ or $-1$, constant in each unipolar zone outlined by the neutral lines reconstructed from the available H$_\alpha$ synoptic maps).

We note that the time when a 
{given $P$}
%polar field measure 
peaks may differ from
the sunspot cycle minimum by several years. Usually, the 
{$P$ proxies}
%polar field 
peak one to two years before the sunspot cycle minimum.  Hence, evaluating
{$P$s}
%polar field measures 
at solar minimum yields somewhat different values
for these correlations, which are nevertheless found to be similar
to the values 
{for peak $P$s}
shown in \Fig{fig:obscorr}. 
{When evaluated at sunspot minimum,} the correlation
values are, respectively for the northern and southern
hemisphere, $r=0.86$ ($p=0.14$) and $r=0.86$ ($p=0.14$) for the WSO polar 
field data, and $r=0.60$ ($p=0.06$) and $r=0.66$ ($p=0.04$) for the {MWO} 
facular data ---comparable to those reported in \cite{muno12} from the
same facular data.

%; see \Tab{tab:precursor}.

\begin{figure}
\centering
\includegraphics[scale=0.36]{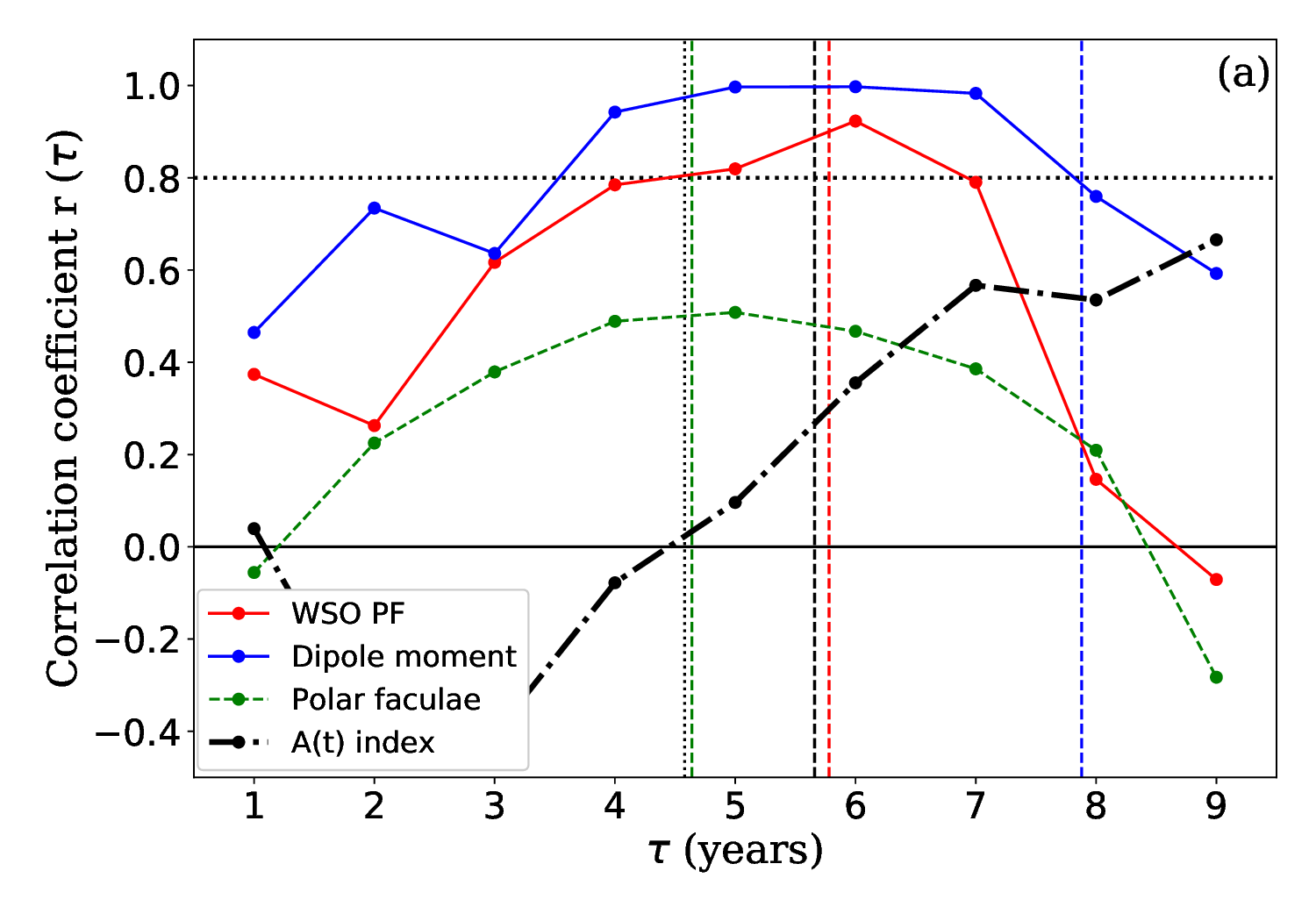}
\includegraphics[scale=0.36]{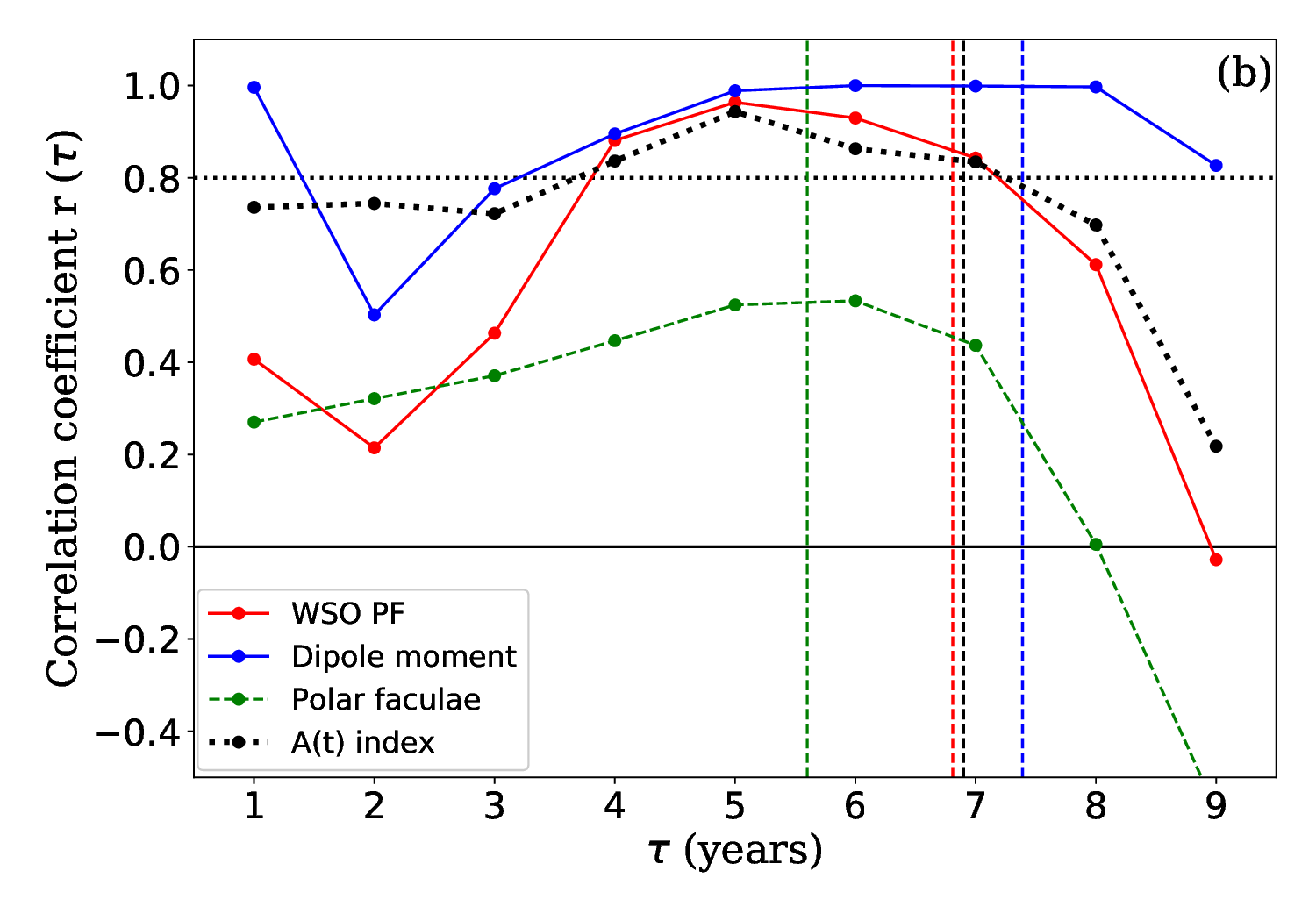}
\caption{Pearson's correlation coefficient $ = {\rm Corr} [T(t_{\rm max}), P(t)]$ between the peak sunspot area $T$ at the maximum of the next cycle and polar field measures $P(t)$ as a function of time 
(a) measured backward from the cycle maximum: $t=t_{\rm max} - \tau$, (b) measured forward from the reversal of $P$: $t_{\rm rev} + \tau$.
Vertical lines indicate the average positions of other cycle landmarks in the given plot, with the same color coding as the associated curves. (a) Black dots: cycle minimum $T_{{\rm SSA,min},i}$, dashed lines: time shift from the time of the maximum of $|P|$ to next cycle maximum
$T_{{\rm p,max},i-1} \rightarrow T_{{\rm SSA,max},i}$.
(b) Dashed lines: time shift from reversal of $P$ to cycle minimum 
$T_{{\rm p,rev},i-1} \rightarrow T_{{\rm SSA,min},i}$. 
}
\label{fig:obslag}
\end{figure}

{To proceed towards the main purpose of this analysis,}
%Next, we 
we now consider the time dependence of these correlations.
\Fig{fig:obslag} shows the correlation coefficient values against time
$\tau$ (in years) measured from two different solar cycle landmarks:
{backward from} solar maximum and 
{forward from poloidal} 
%polar
field reversal. The curve labeled WSO here
shows the WSO polar field results. 

For WSO and polar facular data the correlations are calculated based on a data
set consisting of the hemispheric values of the polar field vs the
hemispheric amplitude of the next cycle, for both hemispheres in the
same data set. This effectively doubles the sample size.

The curves obtained are not smooth, presumably due to the limited
sample size (limited number of cycles).  Nevertheless, it is apparent
that both panels display a 3--4 year long plateau where correlation
values are consistently high. This is the reason why no significant
difference was found above, between correlation levels for predictors
evaluated at cycle minimum or at the time when they take their peak
values.

The plateaus further suggest that the polar precursor may give reliable predictions at times significantly earlier than solar minimum. Indeed, \Fig{fig:obslag}(a) shows that WSO magnetic field data may correctly predict the amplitude of the next cycle maximum up to 7 years ahead 
---{that is a few} 
years earlier than either the cycle minimum or the time when the polar field measures typically reach their maxima.

While this long temporal range of the polar precursor method is an important finding, for practical applications the precursor needs to be evaluated at a time measured forward from an already known landmark, rather than backward from an as yet unobserved one. Hence, in \Fig{fig:obslag}(b) the correlations are now plotted against time measured forward from the reversal of the given poloidal field measure $P$. It is again apparent that reliable predictions, at a Pearson correlation level above 0.8, may be possible already 4 years after polar reversal, that is on average 3 years before cycle minimum.

Regarding the possible choices of precursors $P$, we see that polar faculae counts are clearly inferior to other 
{$P$s}, 
%considered polar field measures, 
including the $A(t)$ index. Note, however, that omitting Cycle 16 can significantly improve their performance, as discussed above.
The performance of the WSO polar field correlations and of the global dipole moment are comparable. 
%(While it may seem that the polar field reaches
%the correlation level 0.8 year earlier, this is due to the
%fact that the global dipole reverses almost a year earlier than the
%WSO polar field: 5 years after dipole reversal roughly corresponds to
%4 years after WSO polar reversal; cf.~Table~\ref{tab:obsstat}.) 

These results indicate that the temporal range of the polar
precursor method is longer than it is generally thought and that reliable
predictions based on this method can be made 4 years after polar
reversal, which is, on average, nearly 3 years earlier than cycle
minimum and 7 years before the predicted maximum. 
In reality, this temporal range may be even longer and a prediction may be made even earlier just by seeing how rapidly the polar field is growing. As shown in Appendix, there is some correlation between the rise rate of the polar field and the amplitude of the next cycle.

As, however, the
statistical sample size upon which these conclusions are based is
limited, for further support we turn to dynamo models that are capable
to simulate a larger number of solar cycles than 
{those few}
that have been well observed.

\begin{figure}
\centering
\includegraphics[width=0.49\textwidth]{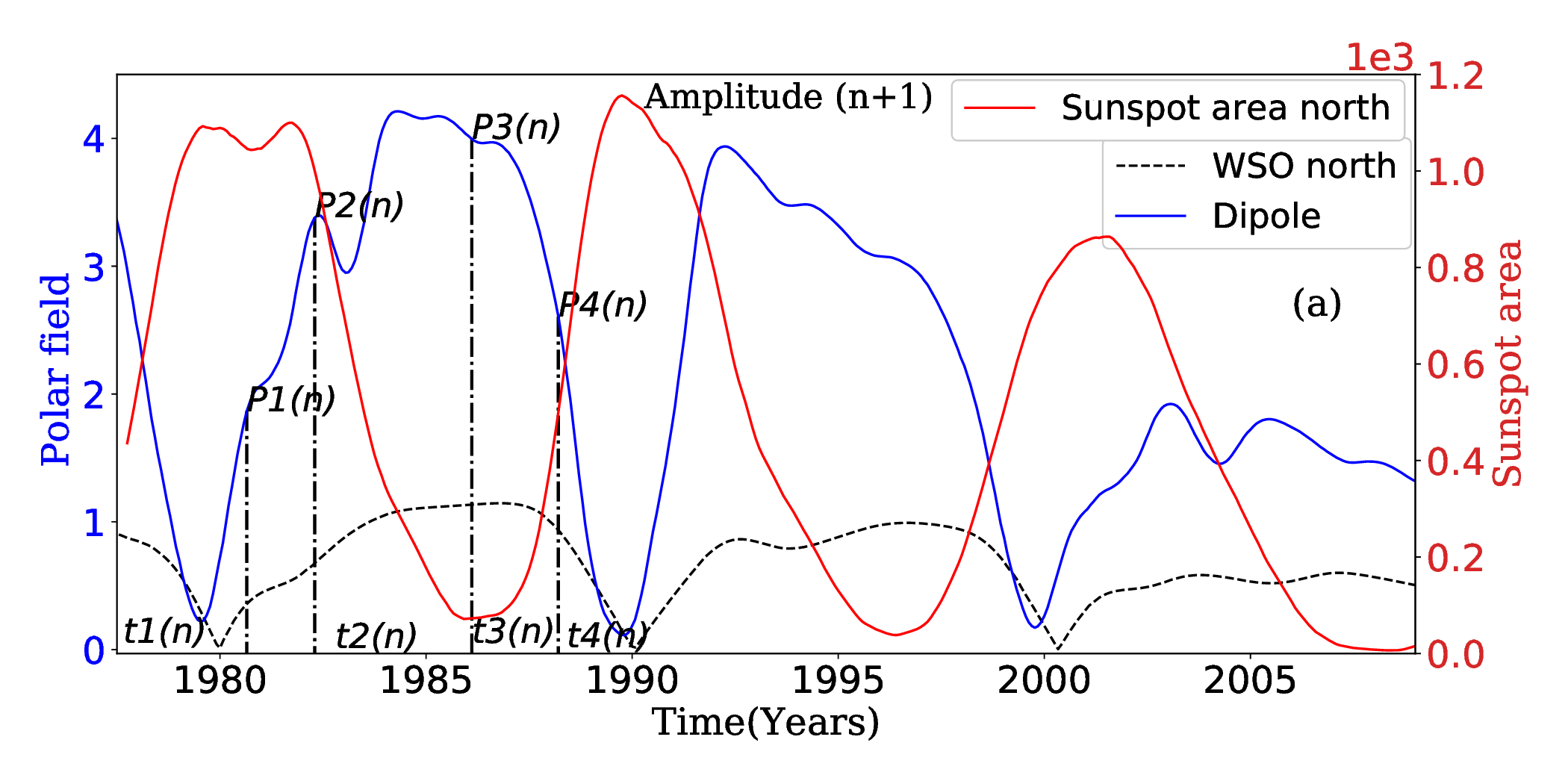}
\includegraphics[width=0.49\textwidth]{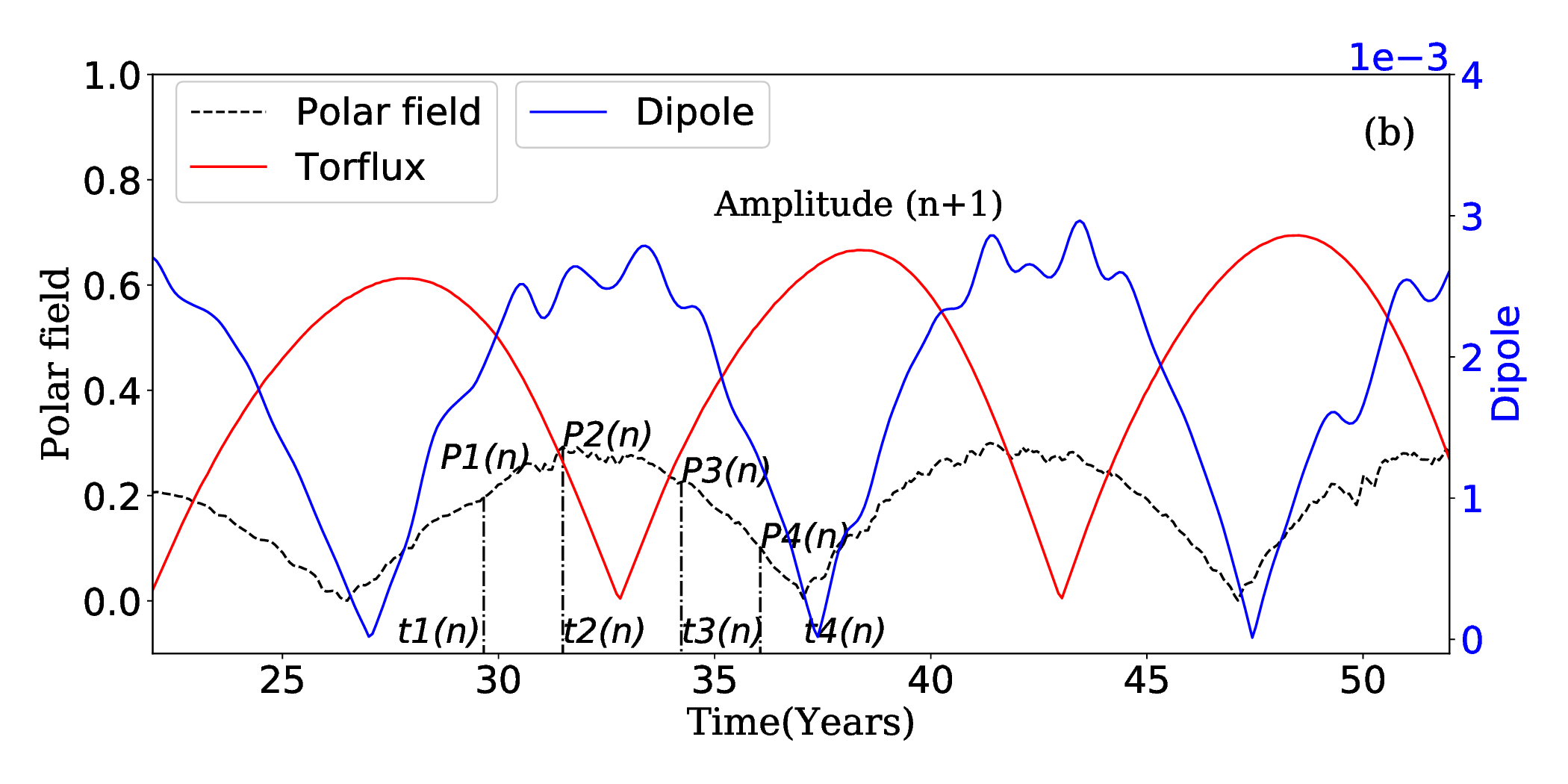}
\includegraphics[width=0.49\textwidth]{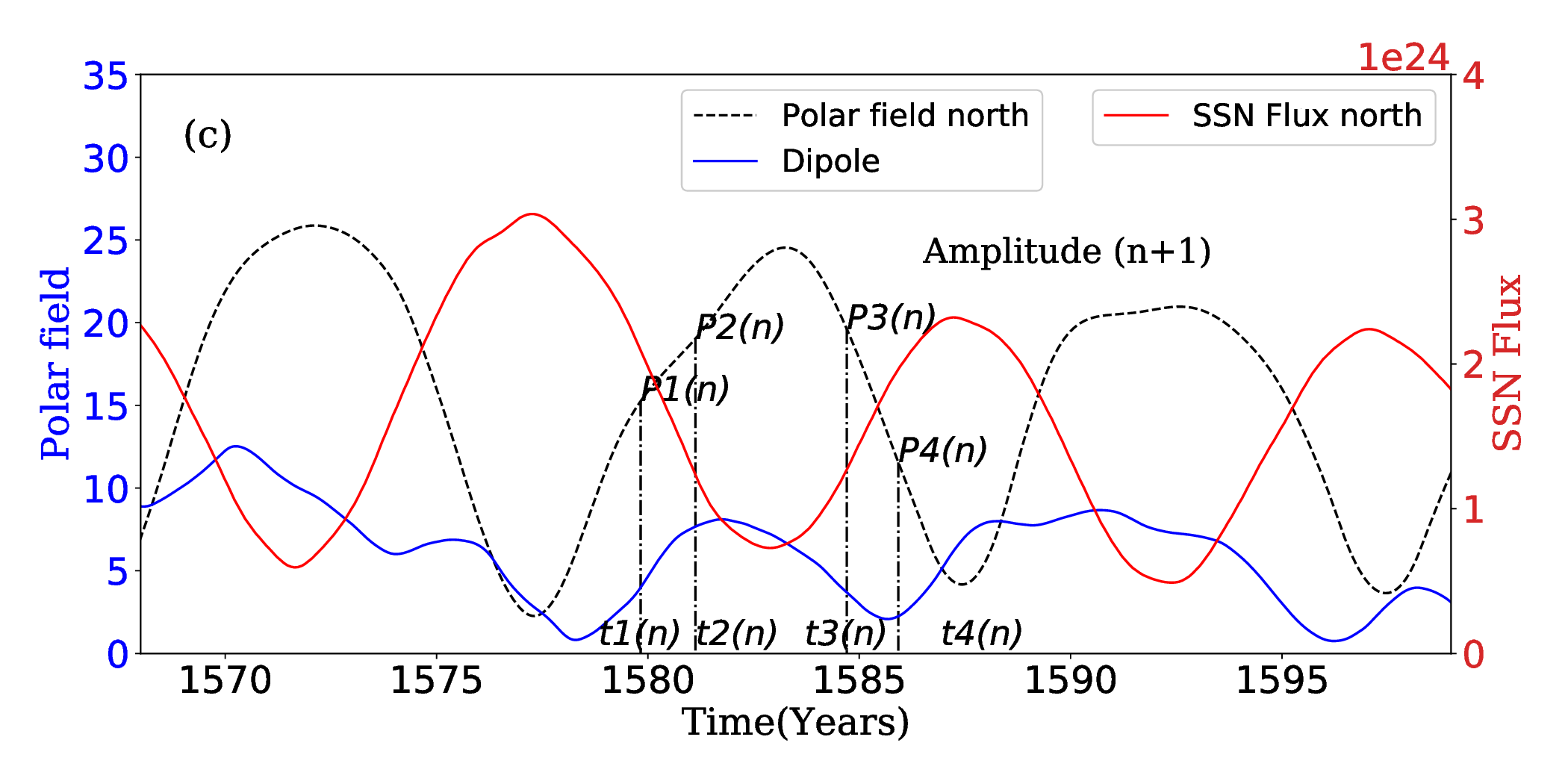}
\includegraphics[width =0.49\textwidth]{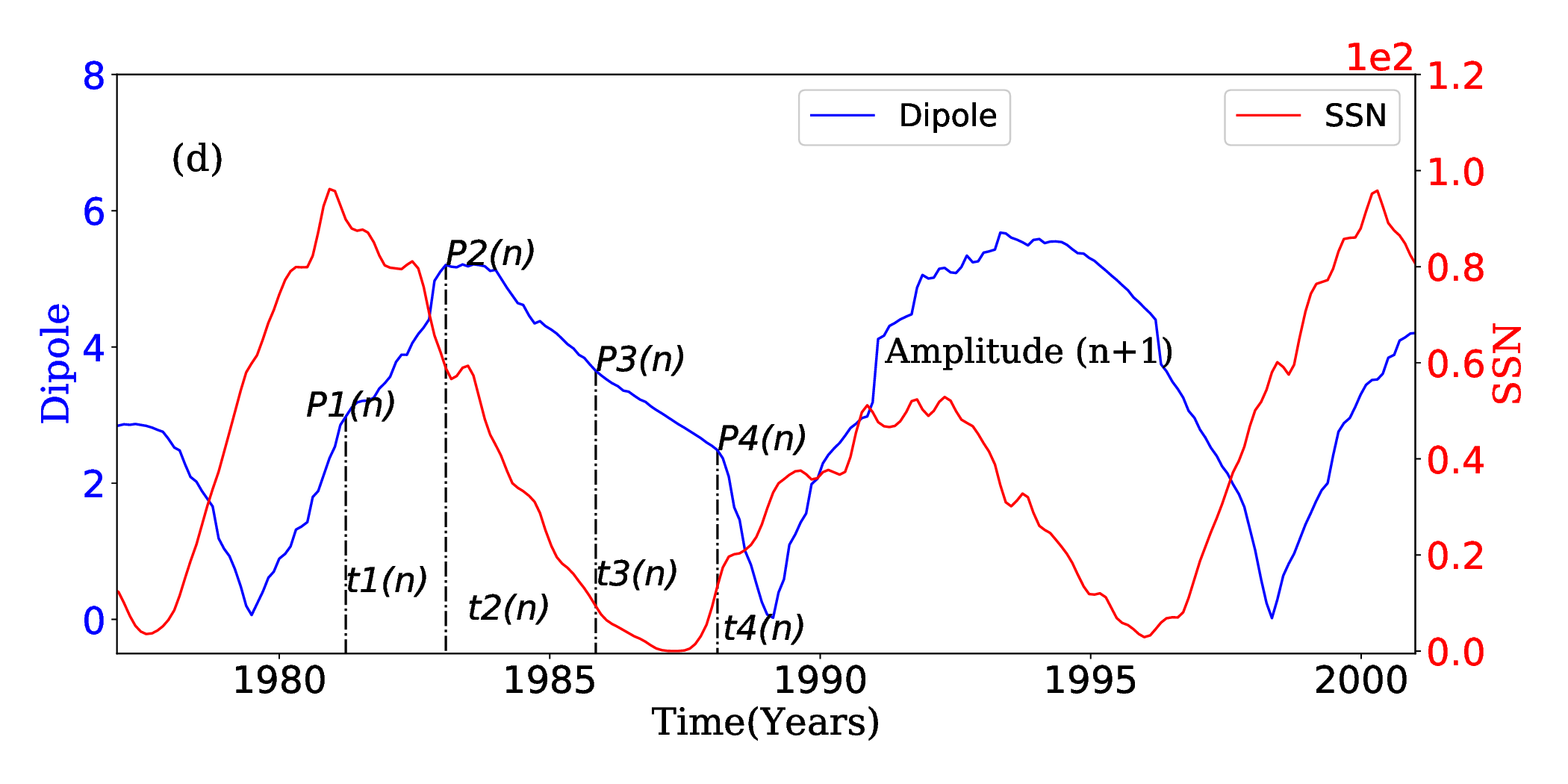} % Melinda
\caption{Time series of some predictors $P$ (blue and black) and predicted activity indicators $T$ (red). {\it (a)} Observations;
{\it (b)} Surya 2DR1 model;
{\it (c)} STABLE 3DR1 model;
{\it (d)} 2$\times$2D-R1 model. 
Some example intervals used for the determination of rise and decay rates (see Appendix) of $P$ are marked.
Rise and decay rates in cycle $n$ are
defined as $(P2(n) - P1(n)) / (t2(n) - t1(n))$ and $(P4(n) - P3(n)) / (t4(n) - t3(n))$, respectively.
}
\label{fig:demo}
\end{figure}

\section{The polar precursor in solar dynamo models} \label{sec:mod}
It is now understood that 
intercycle fluctuations in solar
activity are due to fluctuations in the amplitude of the poloidal
field that serves as a seed for the strong toroidal field in the next
cycle. Fluctuations in the poloidal field, in turn, arise as a
consequence of the stochastic nature of flux emergence  that, upon the
action of surface flux transport processes, ultimately gives rise to
the polar fields. 
{This is the so-called \bl\ mechanism.}
To simulate intercycle variations two approaches are
therefore open: to introduce stochastic variations directly into the
poloidal component of the dynamo equations (\citealt{CD00}, \citealt{KC11},
\citealt{Kitchat+:stochBL}, \citealt{Cameron+:normalform}) or to
explicitly incorporate the stochastic flux emergence process into the
model (\citealt{Yeates+:ARdyn}, \citealt{MD14}, \citealt{LC17},
\citealt{Bhowmik+Nandy}). By calibrating the numerous parameters of these
models it is possible to tune them to resemble the observed solar
cycle in their behaviour, which opens the possibility that
assimilating past observed behaviour into the models and considering
future stochastic variations in an ensemble forecast approach, they may
be used for the prediction of upcoming solar cycles. Some forecasts
for Cycle 25 have indeed been presented using this approach
(\citealt{Bhowmik+Nandy}, \citealt{Labonville+}).

In this section we will use three such models to investigate the
relative merits of different measures 
{$P$} 
of the poloidal field as cycle
precursors and their time span.  All these models are kinematic flux
transport dynamos and thus the  large-scale flows, such as meridional
circulation and differential rotation are imposed as guided by
observations. While the differential rotation profile is the same in
all the models, the meridional circulation is somewhat different.

One model, Surya \citep{CNC04} belongs to the first class mentioned above, with stochastic fluctuations introduced directly to the polar field, while the other two models, STABLE \citep{KM17} and 2$\times$2D \citep{LC17} explicitly incorporate randomized individual active region sources.

Before discussing each model individually, it is instructive to
consider how they fare in reproducing the observed time shifts between
solar cycle landmarks. These time shifts are collected in
Table~\ref{tab:obsstat} (to be compared with the observed values
above), while the time profiles of some of the relevant precursors and
activity indicators are shown in Figure \ref{fig:demo}. A closer
comparison between the model outputs and the observations leads to a
{somewhat} 
disappointing conclusion: despite 
{their respective} 
calibration to 
{specific characteristics of}
solar observations, 
{many of the relative phases and time shifts presented in Table~\ref{tab:obsstat} are not properly reproduced by any of the three models.}
%none of the models considered can acceptably reproduce the relative phases and time shifts of basic solar cycles landmarks

To start with, none of the models 
reproduce enough asymmetry in its 
%the characteristic asymmetric 
cycle profiles: maxima fall at a phase of 0.48--0.52 in contrast to the observed value of 0.42. 
The timing of polar field reversals is 
{also out of phase.}
In Surya 2DR1, the polar field reverses earlier than the dipole moment (Fig.~\ref{fig:demo}b), in contrast to observations. In STABLE 3DR1 the average time of the reversal is too late by more than a year and its timing, especially for the dipole moment, shows very wide haphazard intercycle variations. In the 2$\times$2D model the reversal is about a year earlier than observed and tends to fall in the middle of the rising phase of the cycle. 

The reason for these discrepancies is mainly that the calibration of the models to observations is usually judged by a (quantitative or qualitative) agreement between the observed and modeled shapes of the activity butterfly diagram and of the supersynoptic map of the poloidal field evolution. As the polar areas cover a small fraction of the solar surface, they do not weigh much in such comparisons; and a comparison to temporal profiles of activity is often {not impelled (see also \citealt{Petrovay+Talafha}).}
%These discrepancies between the models and the observations indicate that, for credible cycle predictions, reproducing cycle landmarks should be given more weight when the models are calibrated or optimized to observations.

Facing {these irregularities}
we embark in our study of the performance of polar precursors in dynamo models {with reserve.}
What we are looking for is whether, despite their imperfect optimizations, their simplifications, their physical differences and the different parameter regimes in which they operate, there are any features where the behaviour of possible polar precursors agrees in all the models and observations.

\begin{figure}
\centering
\includegraphics[width=0.49\textwidth]{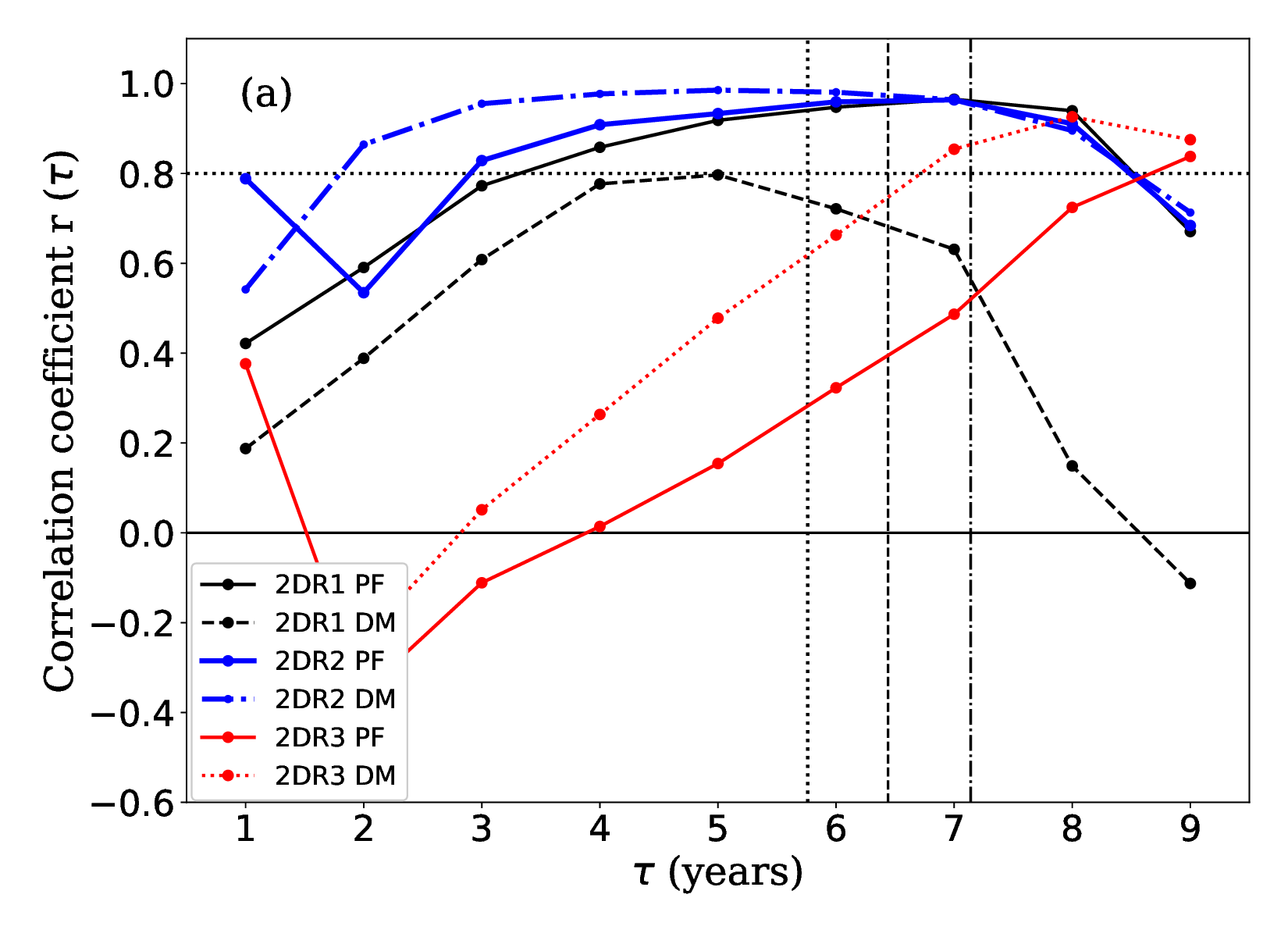}
\includegraphics[width=0.49\textwidth]{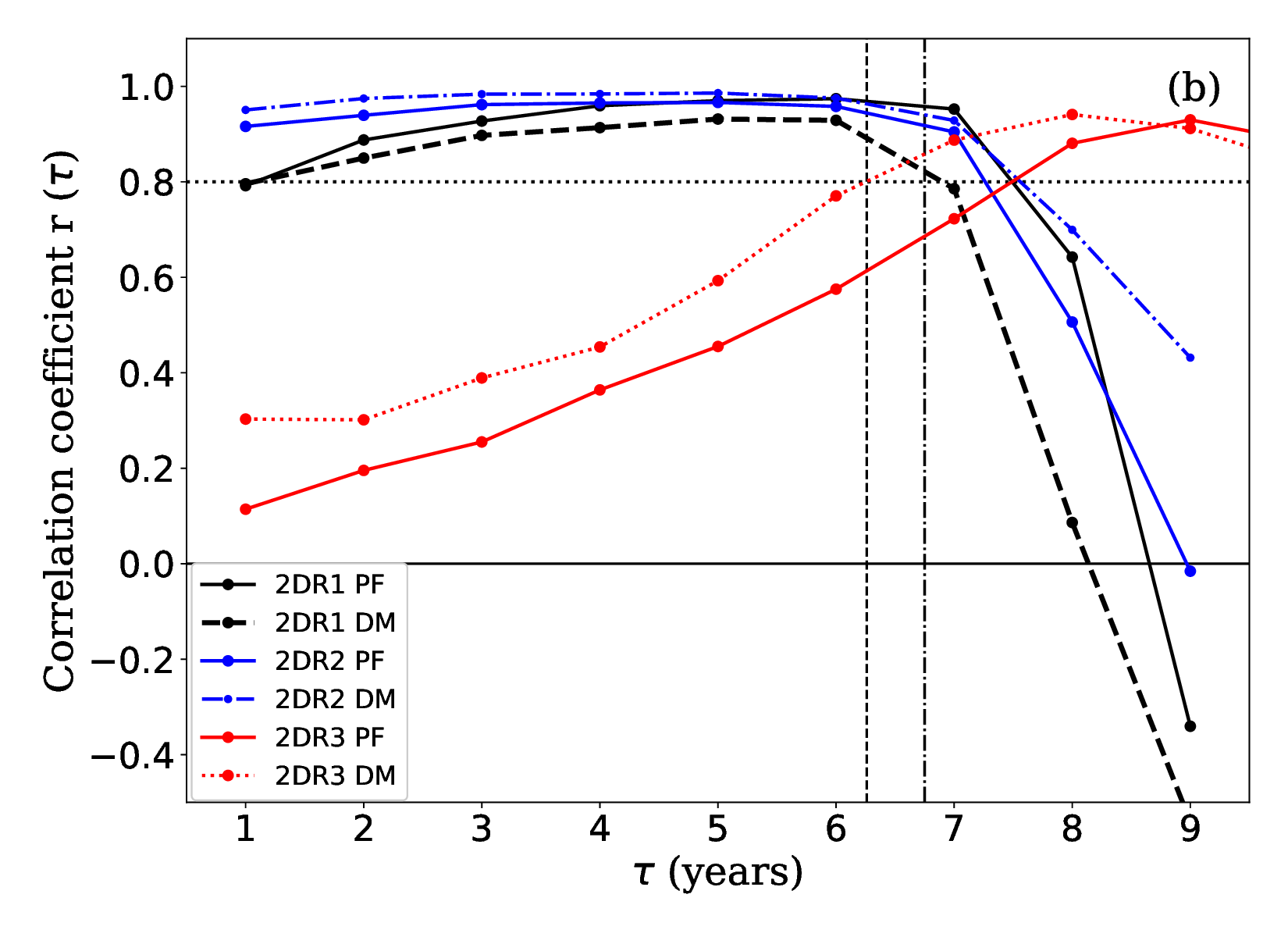}
\caption{Same as Fig.~\ref{fig:obslag}, for different
parameterizations of the Surya dynamo model. PF: polar cap flux;
DM: global dipole moment. Vertical lines mark the mean positions of
cycle landmarks with the same colour coding as the associated curves.
(a) Black dots: cycle minimum $T_{{\rm SSN,min},i}$, dashed lines: time shift from the time of the maximum of $|P|$ to next cycle maximum
$T_{{\rm p,max},i-1} \rightarrow T_{{\rm SSN,max},i}$.
(b) Dashed and dot dashed lines show the time shift from reversal of $P$ to cycle minimum 
$T_{{\rm p,rev},i-1} \rightarrow T_{{\rm SSN,min},i}$, respectively for dipole moment and polar field. 
}
\label{fig:modlag_Surya}
\end{figure}

\subsection{Surya dynamo model}
\label{sec:surya}
Surya is an axisymmetric dynamo model in which  equations for the
poloidal and toroidal fields are solved numerically. Diffusivity for
the poloidal magnetic field in the whole CZ is $3\times10^{12}$~\cmss.
In contrast, below the radius of  $0.975 R_\odot$ the diffusivity for
the toroidal component is reduced to $4\times10^{10}$~\cmss. While
the detailed model was presented in \citet{CNC04}, over the years 
some parameters have been changed in order to make a closer comparison 
with observations \citep{Kar10, Haz15, KMB18}.  For the present study
we use the version of the code that was used in \citet{KC11}.

The basic Surya model tends to produce a regular magnetic cycle. 
However, the variations in the \bl\ process and meridional flow can
produce variations in the magnetic cycle. \citet{KC11} showed that a
combination of  $100\%$ fluctuations with a coherence time of one
month in the \bl\ $\alpha$ and $30\%$ in the meridional flow with a coherence
time of 30 years could produce variations in the solar cycle comparable
to that seen in the observed solar cycle and successfully explain the
Waldmeier effect. We perform a first run with this combination of
fluctuations, labelled 2DR1. We then perform two more
runs for comparison: run 2DR2 is identical to 2DR1
but with the strength of the \bl\ $\alpha$ increased by a factor 2;
and run 2DR3 is identical to 2DR1 but with the diffusivity of the poloidal
field reduced by half.

Activity level $T$ in the model is characterized by the toroidal flux 
at $10^\circ \le$  latitude $\le 30^\circ$ at $r=0.71R_\odot$. For the
precursor $P$, the polar cap flux PF (surface flux poleward of
latitude $75^\circ$) and the global dipole moment DM are considered.
The time variation of the correlation between $P$ and $T$ is displayed
in  Figure~\ref{fig:modlag_Surya}. 

While the low diffusivity run 2DR3 is clearly off, it is seen that the run 2DR2, with elevated poloidal source term, actually produces 
higher correlations
than the reference case 2DR1. In 2DR2 both PF and DM are found to be good predictors of the upcoming cycle amplitude while in 2DR1 for DM this is only true when time is measured from the reversal. Nevertheless, PF and DM are found to be very good predictors of an upcoming solar maximum right from the time of reversal (panel (b)) for both runs, with optimal performance of the predictor reached around cycle minimum.

Some conclusions about the physics of the model versus the real Sun may
also be drawn from these plots. The fact that the correlation in panel (a)
becomes strong for about  $\tau > 4$ years in the first two runs
suggests that the polar field needs at least 4 years to be transported
to the base of the CZ and produce toroidal field for the next cycle.
Keeping in mind the diffusivity  of the poloidal field
($3\times10^{12}$~\cmss), the diffusion time for the poloidal field
to  reach from the surface to the base of CZ is $4.6$~years. Thus, the
delay time in the correlation  in \Fig{fig:modlag_Surya}(a) for runs
2DR1 and 2DR2 is reasonable. As the average time delay from polar
reversal is around 6 years, the correlation is good already very soon
after the time of reversal, in contrast to observations. This suggests
that Surya runs 2DR1 and 2DR2 operate in a somewhat more diffusive regime
than the solar dynamo. In agreement with this, the very
low diffusivity run~2DR3 deviates starkly from observations. There is
a weak negative correlation between 
{$P$ and $T$}
%the toroidal flux and the polar flux 
for $\tau < 7$ years. Then it increases to positive values at
large $\tau$. This clearly shows that in this run,  the model needs at
least about 8 years for the polar flux from the previous cycle to be
transported to the deep CZ to produce strong toroidal flux. The reason
for this delay correlation is the lower diffusivity. In this case, the
diffusivity for the poloidal field is $1.5\times10^{12}$~\cmss\ and thus
the diffusion time is 9.2~years, which is double than that in runs
2DR1 and 2DR2. This explains the strong positive correlation in
Run~2DR3 for $\tau > 8$ years.

\subsection{STABLE dynamo model}
\label{sec:stable}
The STABLE (Surface flux Transport And Babcock--Leighton) dynamo model
was originally developed by Mark Miesch and his colleagues at High
Altitude Observatory \citep{MD14, MT16}. 
Later this model was
significantly improved by \citet{KM17} to make a close connection of
the BMR eruption and evolution on the surface with observations. 
The salient features of this
model are the following. (i) It is a full 3D dynamo model in which
the induction equation is solved over the whole solar CZ.  Unlike the
Surya model, the turbulent diffusivity for all components of magnetic
fields is the same.  The diffusivity has a radial dependent profile
such that in the CZ it has a value of about $10^{12}$~\cmss\
($4.5\times10^{12}$~\cmss\ for $r > 0.956R_\odot$ and
$1.5\times10^{12}$~\cmss\ below) and below the CZ it drops by about
four orders of magnitude. 
The model includes a radial downward magnetic pumping with a speed
20~\mps\ in the top $10\%$ of solar radius to mimic the
asymmetric convection.
(ii) Timing of BMR eruption: This model places
a BMR only when certain conditions are satisfied. First, the magnetic
field at the base of the CZ must exceed a critical field strength.
Second, after the first BMR is produced, the time for the second BMR
will be taken from a log-normal distribution which is obtained from
the observations of BMRs. (iii) Connecting the toroidal field to BMR:
The rate of BMR production is regulated with the strength of the
magnetic field such that when the magnetic field in the base of the CZ
is strong it produces a large number of BMRs. Thus the delay
distribution is regulated by the magnetic field and it is the only
part through which, the toroidal field is linked to the BMRs (and thus
the poloidal field) in this model. (iv) While the magnetic field in
BMR is fixed at 3~kG, the flux is obtained from the observed
distribution. The tilt is obtained from Joy's law with a Gaussian
scatter around it. A magnetic field dependent nonlinear quenching is
imposed in the tilt to saturate the magnetic field in this  dynamo
model. We note that recently some evidence of nonlinear quenching in
the tilt is observed in the BMR data \citep{Jha20}. As shown in
\citet{J20} and \citet{Kar20}, the observed latitude variation of BMRs
provides another nonlinearity to stabilize the magnetic cycles in \bl\
dynamos. In addition to the tilt quenching, this nonlinearity is also
operating in the model to limit the 
amplitudes of the
magnetic cycles.  For further
details of the STABLE dynamo model, we refer the reader to Sections 2
and 4 of \citet{KM17}.

From the simulation presented in \citet{KM17}, we take the runs~B10,
B11 and B13 as listed in their Table~1. We relabelled them here as 3DR1,
3DR2, and 3DR3, respectively. In run~3DR1, a Gaussian scatter of zero
mean and $\sigma =15^\circ$ is included around Joy's law, while in
run~3DR2, the scatter is doubled i.e., $\sigma =30^\circ$.  Finally,
run~3DR3 is the same as run~3DR1, except that the diffusivity below the CZ 
is made the same as that in the CZ ($1.5\times10^{12}$~\cmss).

\begin{figure}
\centering
\includegraphics[width=0.49\textwidth]{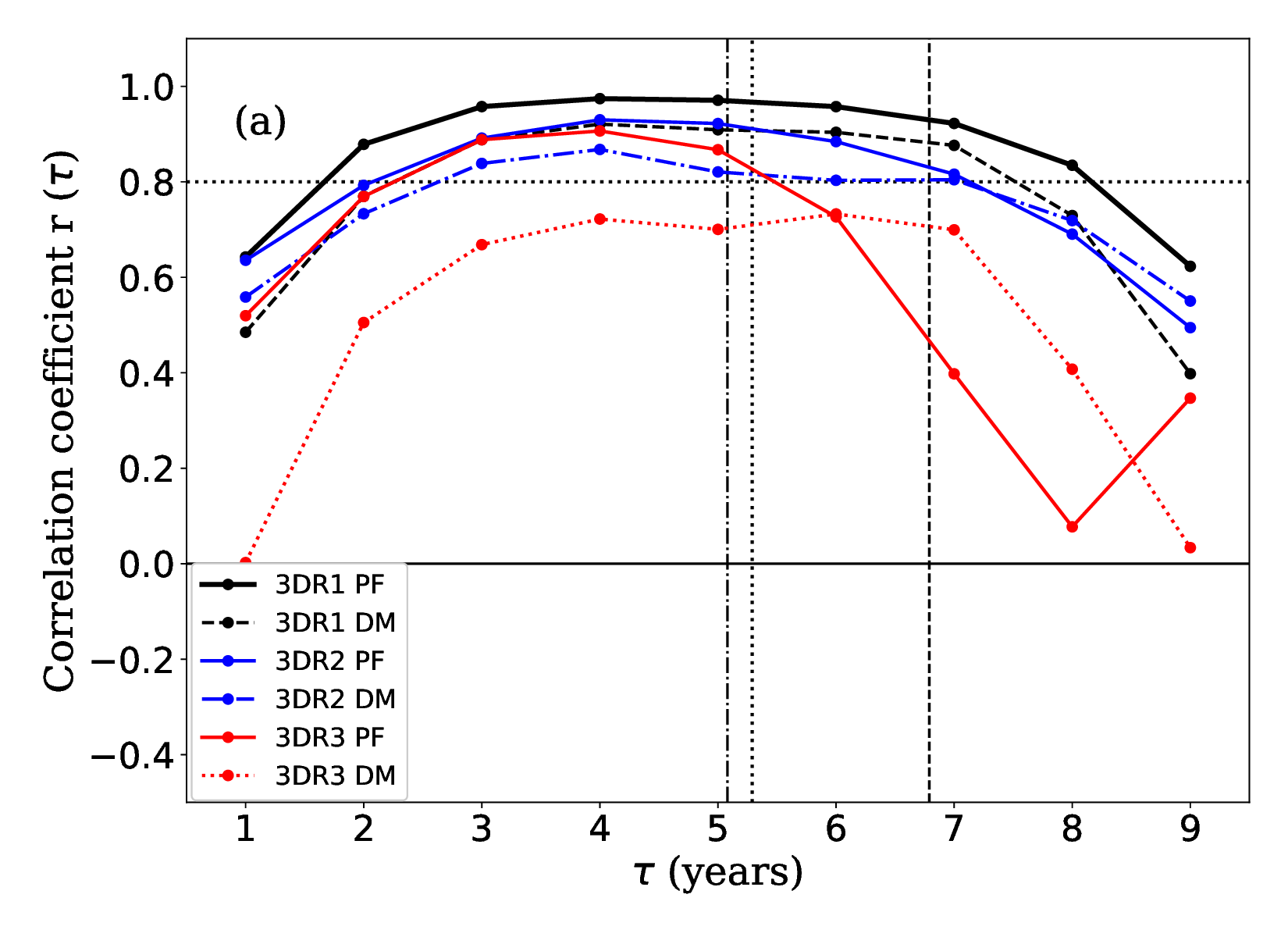}
\includegraphics[width=0.49\textwidth]{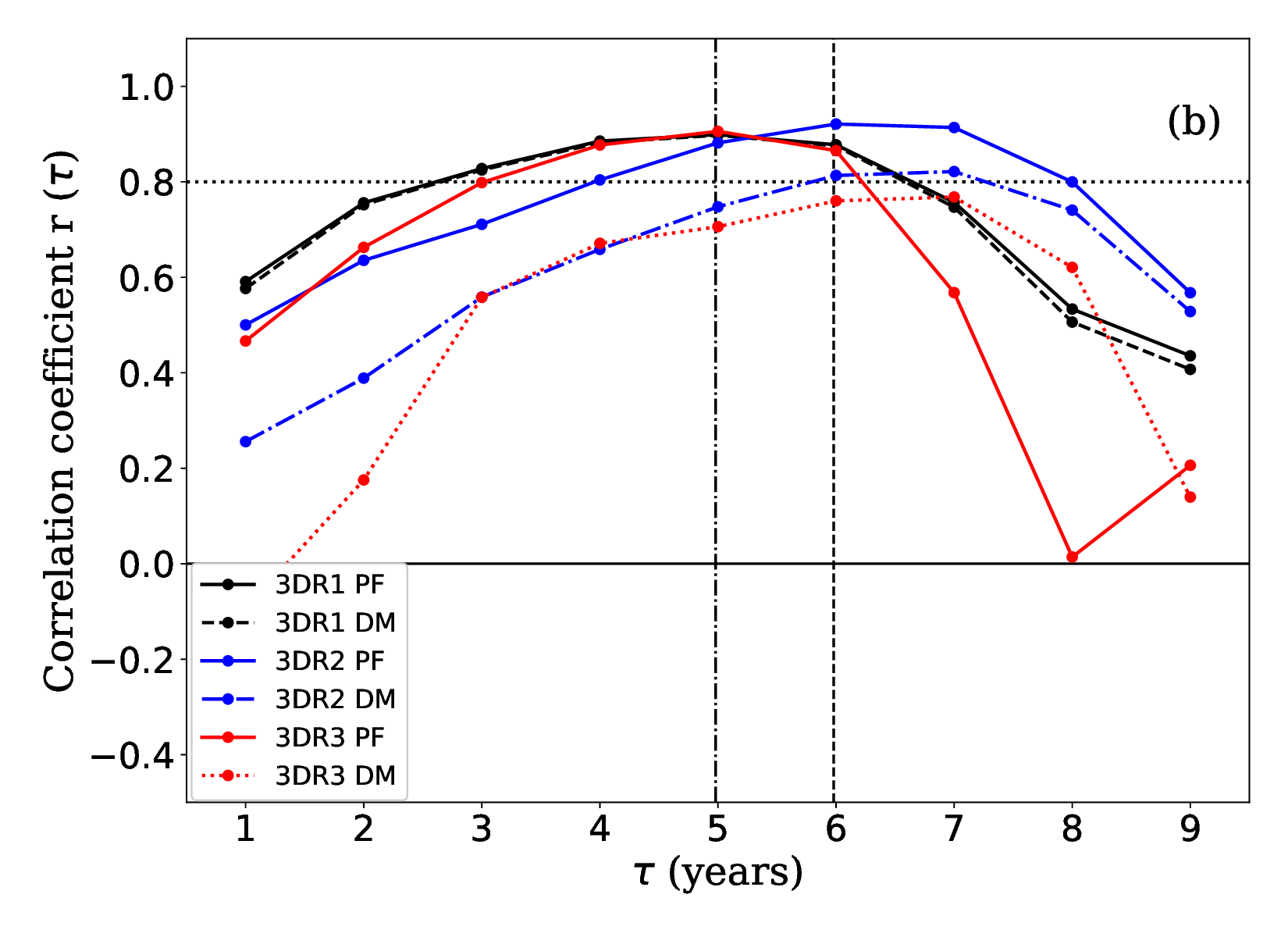}
\caption{Same as Fig.~\ref{fig:modlag_Surya}, for different parameterizations of the STABLE dynamo model.}
\label{fig:modlag_STABLE}
\end{figure}

Activity level $T$ in the model is characterized by the total monthly
flux in emerging BMRs. For the precursor
$P$, the polar cap flux PF (surface flux polewards of latitude
$75^\circ$) and the global dipole moment DM are considered. The time
variation of the correlations between $P$ and $T$ is displayed in
Figure~\ref{fig:modlag_STABLE}. 

In the reference run 3DR1, both PF and DM are found to be very good
precursors of the next cycle amplitude over a broad temporal range
starting 7 years before the predicted maximum or 4 years after polar
reversal. This agrees well with observations. One interesting aspect
of panel (a) is that the correlation exceeds about 0.8 already at
$\tau\sim 2$--$3$ years, implying that the poloidal flux reaches the
deep CZ more quickly in these runs than in the Surya runs.  While the
diffusivity in the STABLE model is comparable to that of the Surya model,
a downward magnetic pumping in the top $10 \%$ of the CZ is an additional
agent for transporting the poloidal flux from the surface to the deep CZ.
This downward pumping may help to produce a strong correlation in
panel~(a) even at small $\tau$ 
{(panel~(b) at large $\tau$)}. 

Run 3DR2, with increased stochasticity, produces somewhat lower correlations, as expected. On the other hand, in panel (a) agreement with observations is improved for larger values of $\tau$ (more rapid fall for $\tau>7$ years). 

In Run~3DR3, with an increased diffusivity below the CZ, the time span
of the correlation shortens to less than about 5 years in panel
(a); nevertheless, in panel (b) a good correlation is still found 4--6
years after 
{$P$}
%polar
reversal. This is related to a significant change in the time shifts between the cycle landmarks (later reversal).
Note that in this case, the DM performs significantly worse as a precursor than the PF.

\subsection{$2\times2$D dynamo model}\label{sec:2by2Dmod} 
The 2$\times$2D hybrid solar dynamo model of \citet{LC17} is also described in detail in several recent publications \citep[e.g.][]{Nagy17}; hence only its most salient features will be summarized here. 
The model is built by coupling 
of a 2D surface flux transport (SFT) module and a 2D axisymmetric flux transport (FTD) module that run concurrently. 
Through the evolving distribution of its internal toroidal field, the FTD module provides the new BMR emergences required by the SFT. In parallel, the SFT module takes care of the spatiotemporal evolution of the surface radial magnetic field and the buildup of the global dipole moment. The zonally averaged surface magnetic field is continuously fed to the internal FTD module via its outer boundary condition.
The probability of new emergences per unit time
and per latitudinal coordinate is set through an emergence function determined by the strength and spatial distribution of the internal magnetic field. This emergence function contains a lower threshold as well, below which the seed magnetic field is considered too small to be amplified and trigger new emergences.
When a new BMR emerges in the model, 
its flux, tilt and separation
are drawn from statistical distributions built from observations of solar cycle 21 \citep{WS89}: 
a log-normal distribution for the flux, a Joy's law for the tilt with Gaussian scatter dependent on the flux, and similarly for the bipole separation.
%\addM{The differential rotation profile used in this model is calibrated helioseismically \citep{Charbonneau1999ApJ}, while the meridional flow profile is given as the modified form of \citet{vanBallegooijen1988}.}
The only amplitude-limiting nonlinearity in the model is a tilt quenching 
that depends on the amplitude of the source toroidal field. 
The model was calibrated to observed butterfly diagrams and surface 
supersynoptic maps as described in 
\citet{LCC15} and \citet{LC17}.
The optimal surface diffusivity was found to be $6\times 10^{12}\;{\rm cm}^2\;{\rm s}^{-1}$ 
and the CZ diffusivity $1\times 10^{12}\;{\rm cm}^2\;{\rm s}^{-1}$, with an
overall profile given by the method of \citet{DC99}. Further parameters of the model are summarized in Table 1 of \citet{LC17}.
%In this model the key ingredient of cycle to cycle variation is the tilt scatter of magnetic regions. 
%On the other hand, motivated by the work of \citet{Das10}, the amplitude limitation happens via tilt-quenching which is the only nonliearity in the $2\times2$D model. ... 
We use this optimized setup {for the run 2$\times$2D-R1 described here. A run 2$\times$2D-R2, at half the tilt scatter, and a run 2$\times$2D-R3, with tilt and separation scatter entirely turned off, are also analysed for comparison.
In all three cases, sequences of 260 synthetic}
cycles are used.

\begin{figure}
\centering
\includegraphics[width = .49\textwidth]{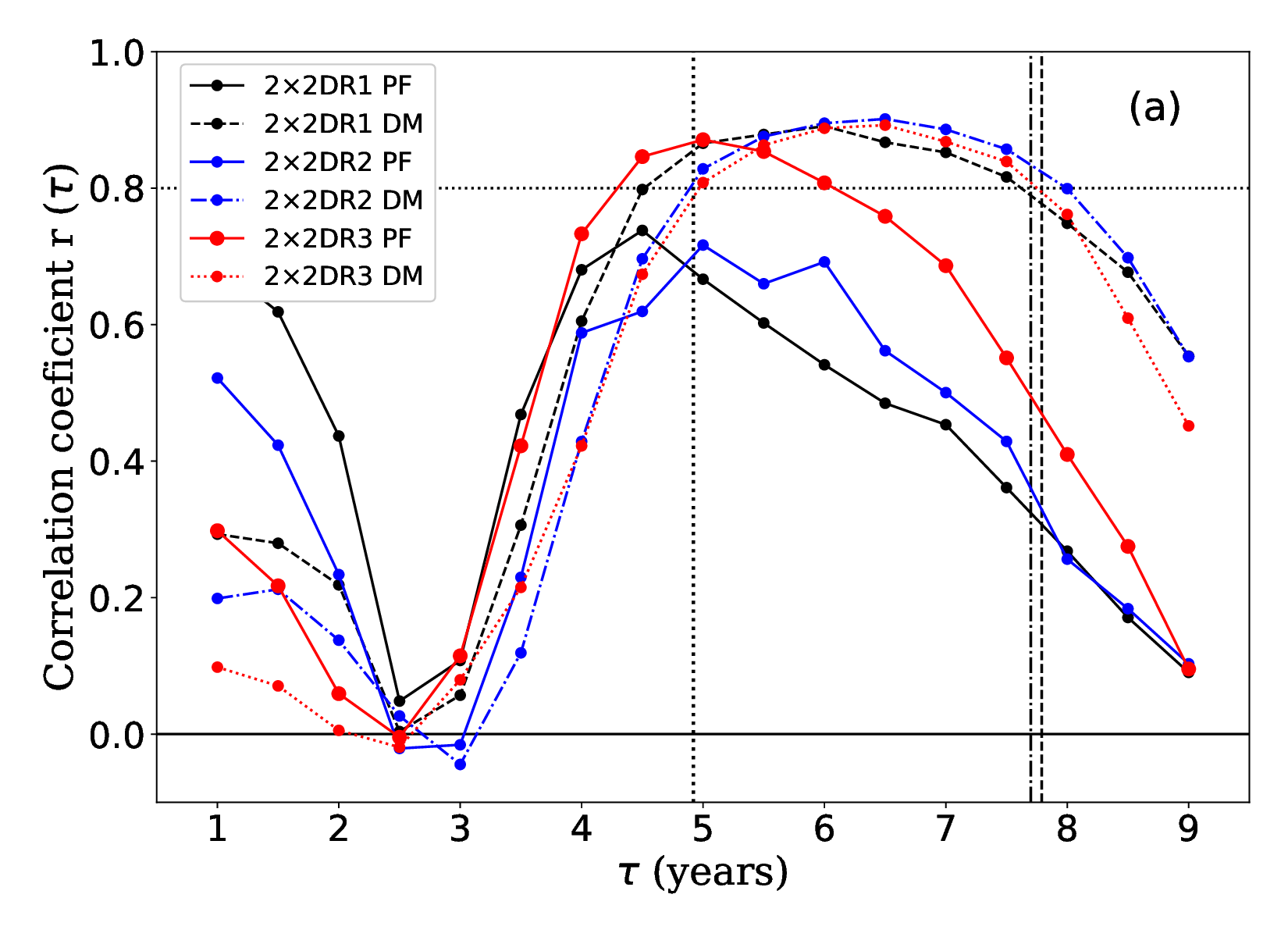} % Melinda
\includegraphics[width = 0.49\textwidth]{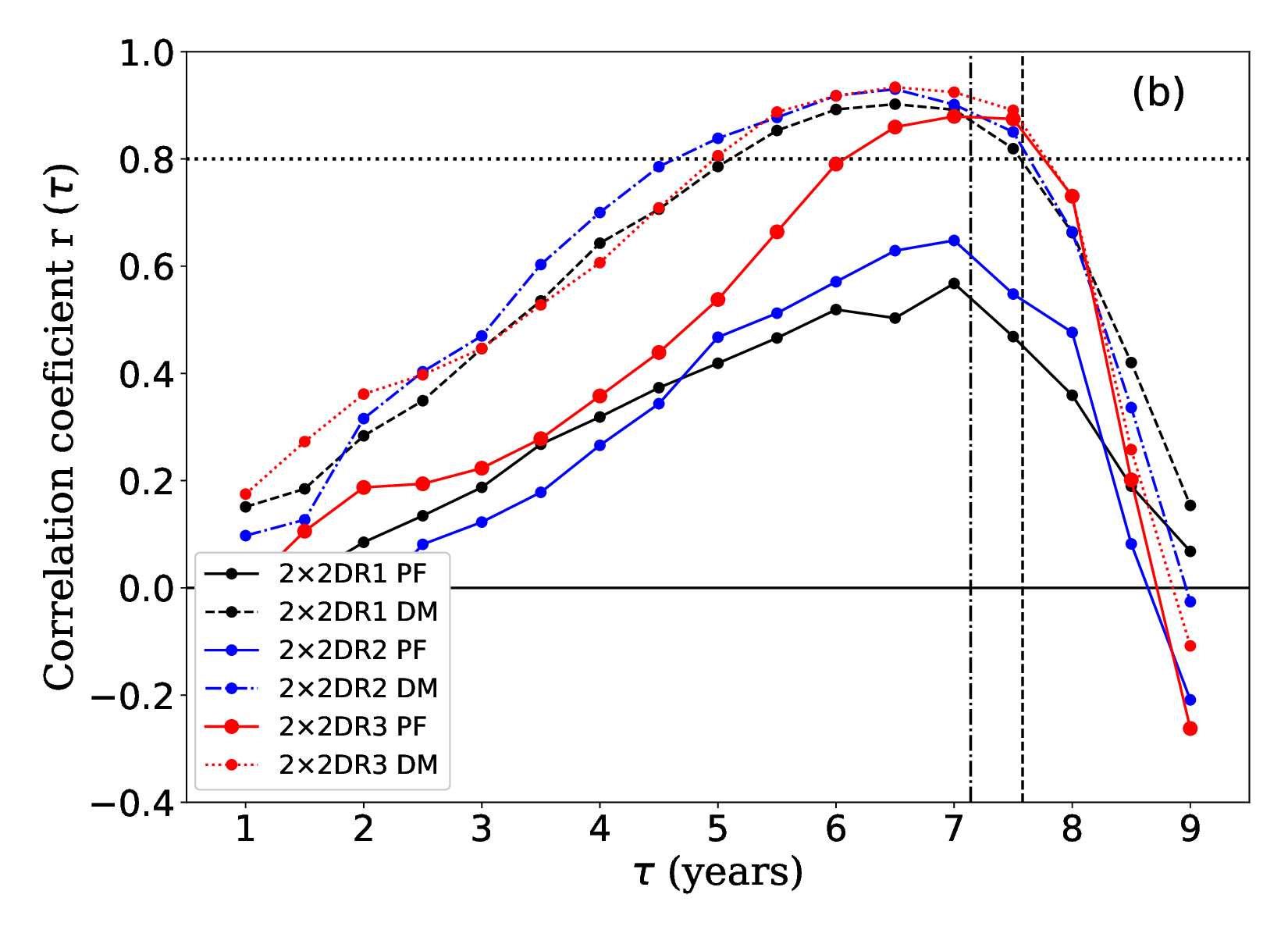} % Melinda
\caption{Same as Fig.~\ref{fig:modlag_Surya}, for the 2$\times$2D dynamo model.}
\label{fig:modlag_2x2D}
\end{figure}

Activity level $T$ in the model is characterized by the {total monthly number of}
emerging BMRs. 
For the precursor
$P$ the polar cap flux PF (surface flux polewards of latitude
$70^\circ$) and the global dipole moment DM are considered. 
The time variation of the correlation between $P$ and $T$ is 
displayed in Figure~\ref{fig:modlag_2x2D}. 

It is noteworthy that in the reference case (R1) only the global dipole moment performs truly well as a precursor. Reducing stochasticity, as in runs R2 and R3, improves the predictive skill of the polar fields, but the dipole moment still remains a better predictor, especially at early times. Correlations with the hemispherically averaged value of the polar field (not shown here) were found to follow a similar trend, so a difference in hemispherical coupling between the model and the real Sun does not explain the poorer performance of the polar fields with respect to the dipole moment. The better predictive skill of the dipole moment must then be related to the importance of fields at lower latitudes, possibly related to the overly dense concentration of the surface flux near the poles, due to the use of a high-latitude peaking latitudinal flow in the last optimized version of the 2$\times$2D model. This incoherence between the FTD and SFT optimizations was noticed in \citet{LC17} and still needs revisiting.
%The return part of the meridional flow (minus the effects of diffusion) may then carry more efficiently the fields located at slightly lower latitudes --which are accounted in the dipole moment calculation-- then those directly near the poles.}
%KP[I would not go into such speculation... It might also be that lowlat. fields directly diffuse down to the bottom etc.]

In the DM correlations, a short plateau is again seen in panel (a), indicating that the DM is a good precursor of the next cycle when evaluated 5 to 8 years before the next cycle maximum (roughly between DM peak and cycle minimum). This agrees well with observations, the extra year in temporal range attributable to the later peak amplitude (near symmetric cycle shape).

Similarly, in panel (b) the DM is found to be a good precursor of the next cycle amplitude 5 to 7 years following its reversal. It should be taken into account that the reversal in this model is more than a year earlier than observed; hence the cycle phase when acceptable predictions become possible is comparable to the observed range 
(which was acceptable for $\tau\gtrsim4$~years after polar field reversal).
Note also that the low correlations in the early years after DM reversal (panel (b)) are also coherent with observations. Before $\tau\approx5$~years, not enough flux has reached the high latitudes to determine the faith of the next cycle, in the observed Sun and especially in this highly stochastic model.

As for the rising rate of polar fields with respect to the amplitude of the next cycle, results presented 
in the Appendix; \Tab{tab:modrrate} indicate that all models show some 
{medium to high}
%correct 
correlations,
%between the rise rate of the polar field and the amplitude of the next cycle as shown in the Appendix; \Tab{tab:modrrate},
{which are similar or higher than observations for the 2D and 3D models, but lower for the 2$\times$2D model.}
The correlations of the decay rate with amplitude presented in the Appendix; \Tab{tab:moddecay} show 
{high values similar to observations for the polar fields (PF) of the 2D and 3D models, and for the dipole moment (DM) of the 2$\times$2D model.}

%%%%%%%%%%%%%%%%%%%%%%%%%%%%%%%%%%%%%%%%%%%%%%%%%%%%%%%%%%%%%%%%%%

\section{Conclusions}
\label{sec:conc}

We have presented an extensive performance analysis of various
measures of the amplitude of the solar poloidal field as precursors
of an upcoming solar cycle, based on both observational data and on
the outputs of various existing flux transport dynamo models. We
calculated correlation coefficients of the predictors with the next
cycle amplitude as a function of time measured from several solar
cycle landmarks. Setting a Pearson correlation level of $r= 0.8$ as a
lower limit for acceptable predictions, observations indicate that the
earliest time when the polar predictor can be safely used is 4 years
after polar field reversal. This is typically 2--3 years before solar
minimum and about 7 years before the predicted maximum, considerably
extending the temporal scope of the polar precursor method.
{Some correlation between the rise rate of the polar field with the next cycle amplitude gives an indication that the temporal scope of the precursor may be even longer in the Sun.}
From the observational record, the polar magnetic field, $A(t)$ index
and global dipole moment are found to perform roughly equally well.

As the statistical sample upon which these conclusions are based is limited to a few well observed solar cycles, for further support of these findings we turned to dynamo models calibrated to realistically represent the observed solar cycle. The three models considered differ greatly in terms of their physical background and in the parameter regime in which they operate. They all involve significant simplifications and, 
unfortunately, all show obvious offsets in their relative phases and timings of characteristic solar cycle landmarks.
%despite being to some extent calibrated to solar observations, we found that none of them satisfactorily reproduces the observed relative timings of major solar cycle landmarks.

Nevertheless, 
all three models being self-consistent \bl\ type, they do produce the strong correlations expected between the poloidal field of an ongoing cycle and the following cycle amplitude.
The stronger the correlations, the more deterministic is the model.
In this sense, the Surya model allows for very early predictions, meaning that the stochastic emergence process plays only little role in the late phases of the cycle; at the other end the highly stochastic 2$\times$2D model is much more restrictive on its predictive window.
The real Sun likely lies somewhere in that interval.
%The process of flux emergence on the real Sun is surely highly stochastic, but still the correlations offer stronger determinism than one might expect.
Thus, where all the models agree, there is a good indicator of some robust feature.
Our analysis of the precursor value and temporal range of polar cap flux and dipole moment in the models has shown that our main conclusion regarding the possibility of using the polar field or the dipole moment as a precursor 4 years after polar reversal holds in all of them.
This supports the robustness of our main observational finding.

The temporal sequence and intervals between cycle landmarks are known
to correlate with the cycle amplitude (the Waldmeier effect being an
obvious example). This raises the possibility that for cycles stronger
or weaker than average the time horizon $\tau$ for which reliable
prediction are possible at $t=t_{p,rev}+\tau$ may show a systematic
deviation from the typical value of 4 years deduced here.
This issue could in principle be addressed by splitting the sample
of cycles into subsamples according to strength and looking for
systematic differences. The shortness of the observational data set,
however, does not allow statistically meaningful inferences from such
small samples. In the case of models, sample size restrictions do not
arise but, as discussed in Section 3 above, the temporal sequence and
spacing of cycle landmarks in these models do not represent observed
solar cycles faithfully enough to give credit to inferences regarding
such finer points as a possible amplitude dependence of the prediction
horizon. Thus, it is at present not possible to exclude the possibility of an amplitude dependence in the result.

To illustrate our main conclusion,

we use WSO polar field measurements in March 2017 (4 years after polar reversal in Cycle 24) to predict the amplitude of the upcoming Cycle 25. For this purpose, linear regressions between the smoothed WSO polar field and peak sunspot area are used. Using hemispherically separated 
data we arrive at a 
prediction\footnote{We note that for the hemispheric prediction, a single regression relation is obtained from the lumped data of two hemispheres and then this relation is used to make the prediction of Cycle 25 from the polar field in the respective hemisphere.} 
of $399.45\pm 105.05$ (North) and $744.38\pm 137.31$ (South), while using $(B_N-B_S)/2$ as a predictor the total peak sunspot area is predicted as $553.29\pm 166.63$. This translates to a peak sunspot ``number'' 
of $120\pm 25$, based on a linear regression ($\mathrm{SSN} = 0.152 \rm{SSA} + 36.565$) between the sunspot number (Version 2{\footnote{\url{http://www.sidc.be/silso/datafiles}}})  vs. sunspot area.

Cycle 25 is now officially considered by SILSO to have started in
December 2019. Using the predictor values at this epoch, a similar
linear regression analysis yields the following updated predictions
for the hemispheric peak sunspot areas: $663.71\pm 91.62$ (North), $523.94\pm 82.27$ (South).
Again based on $(B_N-B_S)/2$ the total peak sunspot area is
predicted as $589.11\pm 21.84$.
(This translates to a peak sunspot ``number'' 
of $126\pm 3$). We note that in these regression analyses, we got only three
complete cycles; the WSO data for Cycle 20 was not complete. 

For the calculation of the prediction error,  we have take the standard deviations in the slope and intercept of the linear regression based on Bayesian probabilistic approach using Python's Pymc3 routine.
We also note that the error in the latter prediction based on the minima of $(B_N-B_S)/2$ is very small because all three data points lie almost on a straight line, however, this did not happen in the prediction based on $(B_N-B_S)/2$ at 4 years after the reversal.

%\blue{\bf{NOTE: In above the regression is based on the peak sunspot area and peak sunspot number during the data we used. We are thinking that instead of using sunspot area, if make the regression relation between the polar field and the peak sunspot number, then we get the predicted value directly in ``number". As we do not have hemispheric sunspot number, we shall do for average polar field only. Here are the results.
%The predicted sunspot number for cycle 25 using linear regression between ``the sunspot number'' and the $(B_N-B_S)/2$ polar field 
%at 4 years after polar reversal in Cycle 24 is $117\pm 12$, 
%while for the polar field at cycle minimum it is $125\pm 3$.}}

It is thus apparent that the prediction made 3 years earlier, that is 4 years
after the polar reversal, already yields a good approximation to the
final forecast. Solar cycle 25 is predicted to be similar to or only
slightly stronger than the previous cycle.

\begin{acknowledgements}

Authors thank the anonymous referee for raising valuable questions which helped to clarified some points.

This research was supported by the Department of Science and
Technology (SERB/DST), India through the Ramanujan Fellowship (project
no SB/S2/RJN-017/2018) and ISRO/RESPOND (project no
ISRO/RES/2/430/19-20),
by the Hungarian National Research, Development and Innovation Fund 
(grant no. NKFI K-128384), by European Union's Horizon 2020
research and innovation programme under grant agreement No. 955620, and by the Fonds de Recherche du Qu\'ebec -- Nature et Technologie (Programme de recherche coll\'egiale).
The collaboration of the authors was facilitated by support from the
International Space Science Institute in ISSI Team 474. 
The Computational support and the resources provided by PARAM Shivay 
Facility under the National Supercomputing Mission, Government of 
India at the Indian Institute of Technology, Varanasi are gratefully 
acknowledged.
\end{acknowledgements}

\bibliographystyle{apj}
\bibliography{paper}% Produces the bibliography via BibTeX.

%%%%%%%%%%%%%%%%%%%%%%%%%%%%%%%%%%%%%%%%%%%%%%%%%%%%%%%%%%%%%%%%%%%%
%%%%%%%%%%%%%%%%%%%%%%%%%%%%%%%%%%%%%%%%%%%%%%%%%%%%%%%%%%%%%%%%%%%%

\clearpage
\appendix
\label{sec:appendix}

\centerline{RISE AND DECAY RATES OF THE POLAR FIELD AS CYCLE PRECURSORS}

\strut

Following a reversal, the observed polar field tends to increase at a
fast rate to reach a plateau lasting until the next solar minimum
(cf.~Fig.~\ref{fig:demo}). This suggests that the level of this
plateau, and hence the predictor value 4 years after reversal might be
anticipated based on the rate of this initial rise (or even the rate
of the decay of the opposite polarity polar field before the
reversal). This might push the temporal scope of the polar precursor
even further back in time. Indeed, \citet{Petrovay+:greenlpred} found
that the rate at which the ``rush to the pole'' feature in coronal
green line supersynoptic maps tends to the poles may be used as a
precursor for the {\it time} of the next cycle maximum. Prompted by
these considerations, here we take a closer look at any possible
correlations between the rise/decay rate of the polar field around
reversal and the amplitude of the next solar maximum.

\begin{table*}[b]
\centering
\caption{Pearson's correlation coefficients ($r$) and the $p$ values computed between the rise rate of the polar field/proxies and the amplitude of the next cycle maximum: observations.
}
\begin{tabular}{lcccccccccc}
\hline
                                        &\multicolumn{2}{c}{Polar faculae} && \multicolumn{2}{c}{WSO polar field} && \multicolumn{1}{c}{A(t) Index}   &&    \multicolumn{1}{c}{DM}\\
\cline{2-3}
\cline{5-6}
\cline{8-8}
\cline{10-10}
Rise rate                               &\multicolumn{2}{c}{$r$ ($p$)} && \multicolumn{2}{c}{$r$ ($p$)} && \multicolumn{1}{c}{$r$ ($p$)}  &&  \multicolumn{1}{c}{$r$ ($p$)}\\
\cline{2-3}
\cline{5-6}
\cline{8-8}
\cline{10-10}
%\hline
computed from                          &  North       &  South        &&  North        &   South       &&   &&  \\
\hline
$t_{{\rm rev}} +1$ to $t_{{\rm rev}}+2$ &  0.57 (0.08) &  0.28 (0.44)  &&  0.34 (0.78)  & $-0.28$ (0.81)&&  0.18 (0.68) &&    0.57 (0.61)\\
\hline
$t_{{\rm rev}} +2$ to $t_{{\rm rev}}+3$ &  0.19 (0.61) &  0.56 (0.09)  &&  0.69 (0.51)  &  0.99 (0.08)  &&  0.64 (0.04) &&    0.67 (0.53) \\
\hline
%$t_{{\rm rev} +3}$ to $t_{{\rm rev}+4}$ &  0.11 (0.75) &  0.69 (0.02) % &&  0.50 (0.67)  &  0.50 (0.67)  &&  0.83 (0.01) &&     0.5 (0.67)\\
%\hline
$t_{{\rm rev} +1}$ to $t_{{\rm rev}}+3$ &  0.47 (0.16) &  0.42 (0.23)  &&  0.95 (0.19)  &  0.01 (0.99)  &&  0.67 (0.07) &&    0.63 (0.56) \\
\hline
%$t_{{\rm rev} +2}$ to $t_{{\rm rev}+4}$ &  0.09 (0.80) &  0.74 (0.01) % &&  0.50 (0.67)  &  0.50 (0.67)  &&  0.64 (0.08) &&    0.5 (0.67) \\
%\hline
$t_{{\rm rev}}$ to $t_{{\rm rev}}+3$    &  0.51 (0.13) &  0.61 (0.06)  &&  0.98 (0.11)  &  0.98 (0.14)  &&  0.77 (0.02) &&     0.48 (0.68)\\
\hline
\hline
\end{tabular}
\label{tab:obsrise}
\tablecomments{Values in the first column are in year and thus the rates are obtained in year$^{-1}$.}
\end{table*}

\begin{figure}
\centering
\includegraphics[scale=0.35]{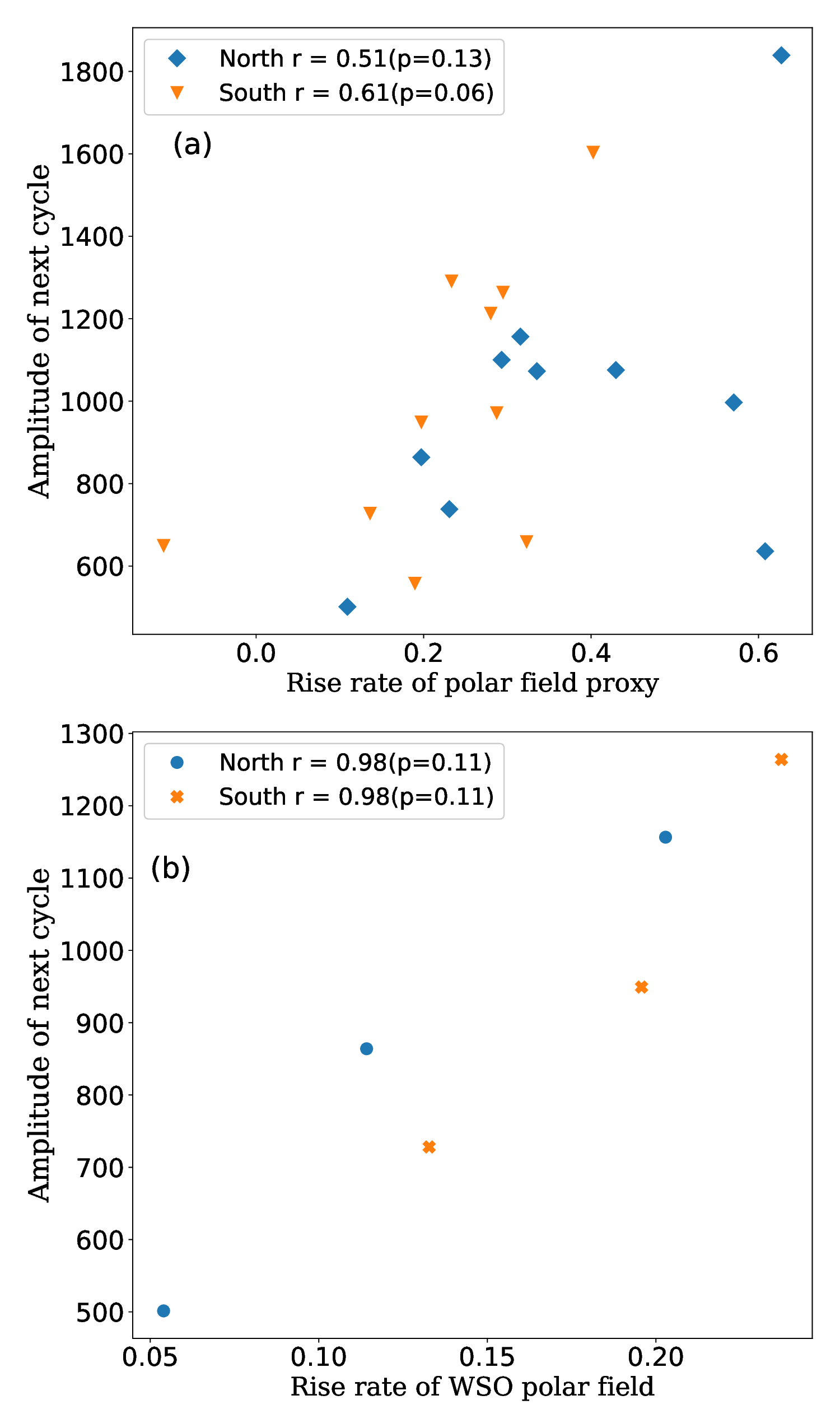}
\includegraphics[scale=0.35]{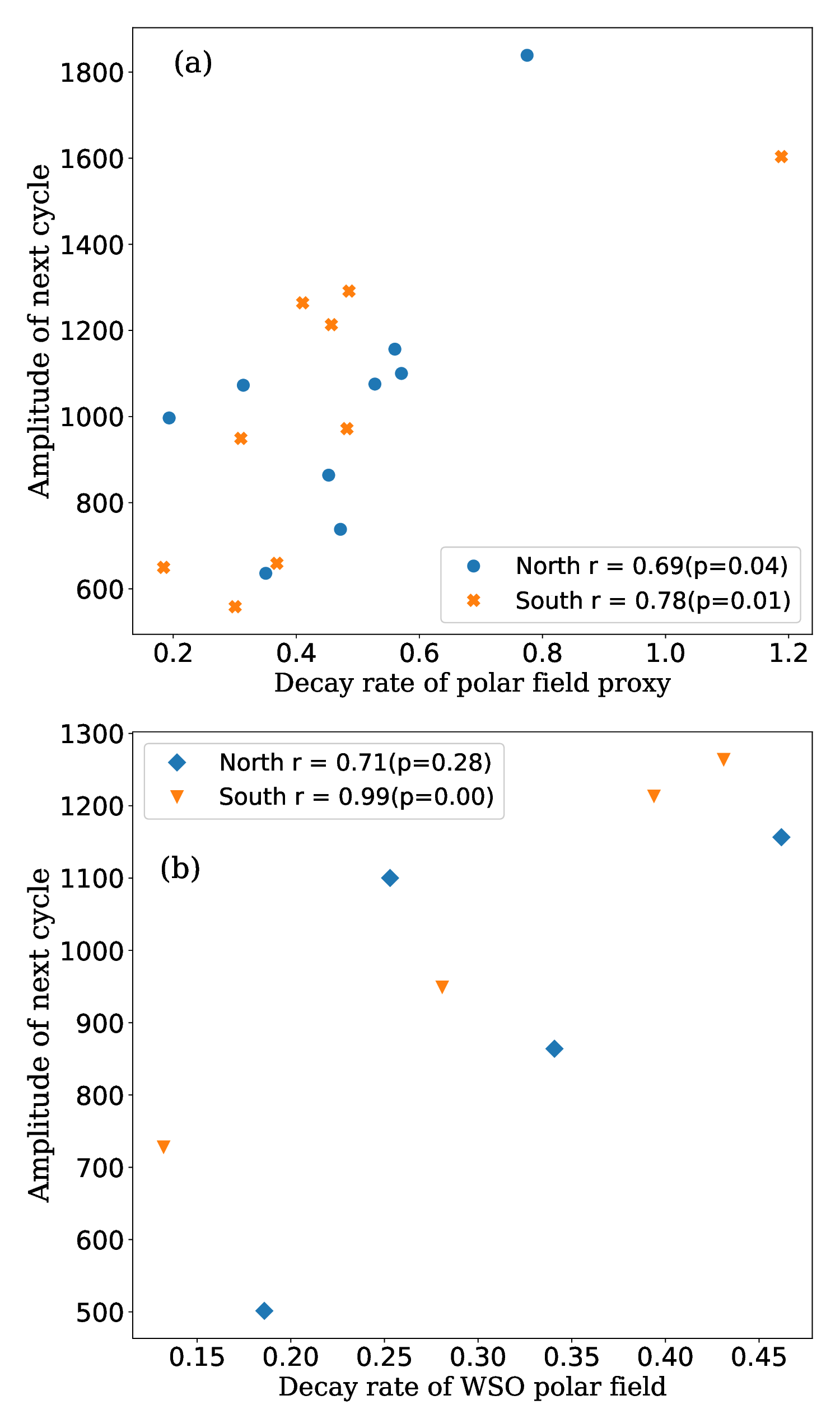}
\caption{Scatter plots between the rise rate (left) or decay rate
(right) a predictor (top: MWO polar faculae count; bottom: WSO polar
field) against the amplitude of the next cycle maximum.
The rates are evaluated from the values taken at $t_{\rm rev}$ and $t_{\rm rev}+3$ (rise) and $t_{{\rm rev} -2}$ and $t_{{\rm rev}-1}$ (decay).
}
\label{fig:obsrise}
\end{figure}

First we consider the rise rate. The rate can be computed in
different ways and at different times in the rising phase of the polar
field. We consider two points at a separation of $\Delta T$ and define
the rate as the difference between the polar fields at these two
points devided by $\Delta T$. This is illustrated in \Fig{fig:demo}.
As the bin size of the polar field proxy data is one year, we cannot
take $\Delta T$ less than one year. \Tab{tab:obsrise} shows
correlation coefficients for the rise rates computed at different
times. For two particular choices of the predictor and the time interval the corresponding scatter plot is shown in \Fig{fig:obsrise} (left). The correlations are generally unconvincing, although the very limited number of data points for WSO measurements does occasionally result in Pearson correlation values close to one there.

Turning now to the decay rate, it is to be noted that the starting and end points of the base interval usually fall in the rise phase of the ongoing cycle. Hence, ``next cycle maximum'' in this case refers to the maximum of the ongoing cycle which is indeed found to be well predicted by the decay rate. This is shown in  \Fig{fig:obsrise} (right), with the correlation coefficients collected in  \Tab{tab:obsdecay}. This correlation, however, is but a natural consequence of the fact that in a stronger cycle high rates of flux emergence are seen already in the rise phase, and the poleward transport of this new flux induces a more rapid decay of the polar fields left from the previous cycle. This result has little practical importance as the correlation values are generally no better than those of the conventional polar precursor method and the temporal scope is shorter.

\begin{table*}
\centering
\caption{Pearson's correlation coefficients ($r$) and the $p$ values computed between the decay rate of the polar field/proxies and the amplitude of the next cycle maximum: observations.}
\begin{tabular}{lcccccccccc}
\hline
\hline
                                        &\multicolumn{2}{c}{Polar faculae} && \multicolumn{2}{c}{WSO polar field}  && \multicolumn{1}{c}{A$(t)$ Index}  &&    \multicolumn{1}{c}{DM}\\
\cline{2-3}
\cline{5-6}
\cline{8-8}
\cline{10-10}
Decay rate                       &\multicolumn{2}{c}{$r$ ($p$)} && \multicolumn{2}{c}{$r$ ($p$)} && \multicolumn{1}{c}{$r$ ($p$)}  && \multicolumn{1}{c}{$r$ ($p$)}\\
\cline{2-3}
\cline{5-6}
\cline{8-8}
\cline{10-10}
%\hline
computed from                             &   North    & South        &&  North      &  South    &&       &&   \\
\hline
$t_{{\rm rev}} -2$ to $t_{{\rm rev}}-1$ &  0.69 (0.04)  &  0.78 (0.01)  && 0.71 (0.28) & 0.99 (0.00)    &&  0.39 (0.37) &&  0.75 (0.24)\\
\hline
$t_{{\rm rev}} -3$ to $t_{{\rm rev}}-2$ &  0.66 (0.05)  &   0.67 (0.05) && 0.10 (0.89)   &  0.54 (0.45)  &&  0.63 (0.13) &&  0.58 (0.42)\\
\hline
$t_{{\rm rev}} -4$ to $t_{{\rm rev}}-3$ &  0.29 (0.45)  &   $-0.04$ (0.90) && $-0.91$ (0.26)    & $-0.02$ (0.98)    &&  0.89 (0.01) &&  0.58 (0.42)\\
\hline
%$t_{{\rm rev} -5}$ to $t_{{\rm rev}-4}$ & $-0.18$ (0.64)~~ 0.15 (0.70) && $-0.5$ (0.67), ~ 0.5 (0.67)\\
$t_{{\rm rev}} -3$ to $t_{{\rm rev}}-1$ &  0.73 (0.02)  &   0.85 (0.01) && 0.62 (0.37)   &  0.83 (0.16)  &&  0.69 (0.16) &&  0.66 (0.33)\\
\hline
$t_{{\rm rev}} -4$ to $t_{{\rm rev}}-2$ &  0.47 (0.20)    &   0.32 (0.40) && $-0.43$ (0.71)  &  0.99 (0.01)  &&  0.94 (0.01)   &&  0.58 (0.42)\\
\hline 
$t_{{\rm rev}} -4$ to $t_{{\rm rev}}-1$ &  0.60 (0.08)  &   0.82 (0.01) && 0.94 (0.21)   &  0.99 (0.03)  &&  0.84 (0.02) &&  0.63 (0.36)\\
\hline
\hline
\end{tabular}
\label{tab:obsdecay}
\tablecomments{Values in the first column are in year and thus the rates are obtained in year$^{-1}$.}
\end{table*}

%\clearpage
\begin{table*}
\centering
\caption{Pearson's correlation coefficients ($r$) and the $p$ values computed between the rise rate of the polar field/proxies and the amplitude of the next cycle maximum: models}

\begin{tabular}{lcccccccccccccccccc cc}
\hline
                                       &\multicolumn{2}{c}{2DR1} &&         \multicolumn{2}{c}{2DR2} &&    \multicolumn{2}{c}{2DR3} &&   \multicolumn{2}{c}{3DR1} &&    \multicolumn{2}{c}{3DR2} &&    \multicolumn{2}{c}{3DR3} && \multicolumn{2}{c}{2$\times$2D-R1}\\
\cline{2-3}
\cline{5-6}
\cline{8-9}
\cline{11-12}
\cline{14-15}
\cline{17-18}
\cline{20-21}
Rise rate                              &\multicolumn{2}{c}{$r$ ($p$)}                                     && \multicolumn{2}{c}{$r$ ($p$)}                                    && \multicolumn{2}{c}{$r$ ($p$)}                                    &&  \multicolumn{2}{c}{$r$ ($p$)}                                    &&  \multicolumn{2}{c}{$r$ ($p$)}                                    &&  \multicolumn{2}{c}{$r$ ($p$)} &&  \multicolumn{2}{c}{$r$ ($p$)}\\
\cline{2-3}
\cline{5-6}
\cline{8-9}
\cline{11-12}
\cline{14-15}
\cline{17-18}
\cline{20-21}
%\hline
computed from                           &   PF        &   DM  &&   PF        &   DM      &&  PF      &   DM     &&  PF      &   DM      &&  PF       &   DM &&  PF   &       DM &&  PF & DM\\
\hline
$t_{{\rm rev}} +1$ to $t_{{\rm rev}}+2$ &   0.74    &  0.61      &&  0.83   & 0.93     &&  0.15  & 0.13    &&  0.76   &  0.69    &&   0.60    & 0.43     &&  0.54 & 0.50   && 0.11  & 0.26  \\
\hline
$t_{{\rm rev}} +2$ to $t_{{\rm rev}}+3$ &   0.57 & 0.27   &&  0.66 & 0.79      &&  0.18  &  0.16    &&  0.78   & 0.57    &&   0.59  &  0.40      &&  0.60 & 0.49   &&  0.11     & 0.28  \\
\hline
%$t_{{\rm rev} +3}$ to $t_{{\rm rev}+4}$ &  0.22  & 0.13   &&  0.10 & %0.69     &&  0.21  & 0.21   &&    0.78   &  0.23   &&   0.60    & %0.22     &&  0.61 & 0.02    &&  0.11     & 0.38  \\
%\hline
$t_{{\rm rev}} +1$ to $t_{{\rm rev}}+3$ &    0.84 & 0.69  &&  0.91 & 0.96      &&  0.24  &  0.28     &&  0.79    & 0.72   &&   0.63   & 0.53      &&  0.61 & 0.62    &&  0.23     & 0.41  \\
\hline
%$t_{{\rm rev} +2}$ to $t_{{\rm rev}+4}$ &    0.61 & 0.29   &&  0.61 & % 0.91     &&  0.33  & 0.27      &&  0.84   & 0.54   &&   0.65   &  %0.32      &&  0.71 & 0.31    &&  0.24     & 0.44  \\
%\hline
$t_{{\rm rev}}+1$ to $t_{{\rm rev}}+4$ &   0.87 & 0.70   &&  0.89 &   0.96    &&  0.37  & 0.37    &&   0.88   & 0.75    &&   0.72  &  0.59     &&  0.74 & 0.60    &&  0.35     & 0.58  \\
\hline
\hline
\end{tabular}
\label{tab:modrrate}
\tablecomments{PF and DM are shorthands for polar field and axial dipole moment, respectively.
Values in the first column are in year and thus the rates are obtained in year$^{-1}$.
}
\end{table*}

For completeness, the corresponding results for the dynamo models are presented in Tables \ref{tab:modrrate} and \ref{tab:moddecay}. These results are in agreement with the observational findings discussed above.

\begin{table*}
\centering
\caption{Pearson's correlation coefficients ($r$) and the $p$ values computed between the decay rate of the polar field/proxies and the amplitude of the next cycle maximum: models
}
\begin{tabular}{clccccccccccccccccc cc}
\hline
                                       &\multicolumn{2}{c}{2DR1} &&         \multicolumn{2}{c}{2DR2} &&    \multicolumn{2}{c}{2DR3} &&   \multicolumn{2}{c}{3DR1} &&    \multicolumn{2}{c}{3DR2} &&    \multicolumn{2}{c}{3DR3} &&    \multicolumn{2}{c}{2$\times$2D-R1}\\
\cline{2-3}
\cline{5-6}
\cline{8-9}
\cline{11-12}
\cline{14-15}
\cline{17-18}
\cline{20-21}
Decay rate                          &\multicolumn{2}{c}{$r$ ($p$)}                                     && \multicolumn{2}{c}{$r$ ($p$)}                                    && \multicolumn{2}{c}{$r$ ($p$)}                                    &&  \multicolumn{2}{c}{$r$ ($p$)}                                    &&  \multicolumn{2}{c}{$r$ ($p$)}                                    &&  \multicolumn{2}{c}{$r$ ($p$)} &&  \multicolumn{2}{c}{$r$ ($p$)}\\
\cline{2-3}
\cline{5-6}
\cline{8-9}
\cline{11-12}
\cline{14-15}
\cline{17-18}
\cline{20-21}
%\hline
computed from                           &   PF        &   DM  &&   PF        &   DM      &&  PF      &   DM     &&  PF      &   DM      &&  PF       &   DM &&  PF   &       DM &&  PF & DM\\
\hline
$t_{{\rm rev}} -2$ to $t_{{\rm rev}} -1$ &   0.88   & 0.89     &&   0.83    &  0.98    &&  0.30 & 0.64     &&  0.83 & 0.30       &&  0.60    & 0.46   &&  0.74  &  0.37  &&  0.52   & 0.83\\
\hline
$t_{{\rm rev}} -3$ to $t_{{\rm rev}} -2$ &   0.81   & 0.65    &&   0.67  & 0.91       &&  0.31 & 0.58    &&  0.90 & 0.46   &&  0.72 &  0.42     &&  0.81  & 0.37   &&  0.03   & 0.48\\
\hline
$t_{{\rm rev}} -4$ to $t_{{\rm rev}} -3$ &   0.41   &  0.02   &&   0.24    & 0.68     &&  0.35 & 0.49    && 0.91  & 0.57   &&  0.80 &  0.34     &&  0.70  & 0.41    &&  0.00  & $-0.03$\\
\hline
$t_{{\rm rev}} -3$ to $t_{{\rm rev}} -1$ &   0.91    & 0.89   &&   0.89   & 0.98      &&  0.34 & 0.64   &&  0.91 & 0.47   &&  0.72 &   0.45    &&  0.82  & 0.42   &&  0.32  & 0.79\\
\hline
$t_{{\rm rev}} -4$ to $t_{{\rm rev}} -2$ &   0.78   & 0.48   &&   0.57   & 0.89       &&  0.42 & 0.68   &&  0.94 & 0.59   &&  0.81 &  0.41     &&  0.81  & 0.46   &&  $-0.04$ & 0.28\\
\hline
$t_{{\rm rev}} -4$ to $t_{{\rm rev}} -1$ &   0.91   & 0.86     &&   0.89  &  0.97     &&  0.42 &  0.71   &&  0.95 & 0.65   &&  0.86 &  0.46     &&  0.78  & 0.42   &&  0.13 &  0.68\\
\hline
\hline
\end{tabular}
\label{tab:moddecay}
\tablecomments{Values in the first column are in year and thus the rates are obtained in year$^{-1}$.}
\end{table*}

\end{document}